\documentclass[12pt]{article}
\usepackage{latexsym,amssymb,amsfonts,amsmath}
\usepackage{mathrsfs}
\usepackage[dvips]{graphicx}

\usepackage{amsthm}
\usepackage{booktabs}
\usepackage{float}
\usepackage{nicefrac}
\usepackage{dsfont}
\usepackage{mathrsfs}
\usepackage{verbatim}
\usepackage{siunitx}
\usepackage{blindtext}
\usepackage{mcite}
\usepackage{tabularray}
\usepackage{slashed}

\newlength{\extraspace}
\newlength{\extraspaces}
\newcommand{\be}{\begin{equation}
\addtolength{\abovedisplayskip}{\extraspaces}
\addtolength{\belowdisplayskip}{\extraspaces}
\addtolength{\abovedisplayshortskip}{\extraspace}
\addtolength{\belowdisplayshortskip}{\extraspace}}
\newcommand{\ee}{\end{equation}}

\newcommand{\bea}{\begin{eqnarray}
\addtolength{\abovedisplayskip}{\extraspaces}
\addtolength{\belowdisplayskip}{\extraspaces}
\addtolength{\abovedisplayshortskip}{\extraspace}
\addtolength{\belowdisplayshortskip}{\extraspace}}
\newcommand{\eea}{\end{eqnarray}}

\DeclareMathOperator{\diag}{diag} 

\newcommand{\Overline}[2][1]{%
 {}\mkern#1mu \overline{\mkern-#1mu #2 \mkern-#1mu}\mkern#1mu {}}

\renewcommand{\Bar}{\Overline[2]}

\newcommand{\rad}{\sqrt}

\begin{document}

\addtolength{\baselineskip}{.8mm}

{\thispagestyle{empty}


\begin{center}
\vspace*{1.0cm}
{\Large\bf Study of the scalar and pseudoscalar meson mass spectrum above the QCD chiral phase transition, using an effective Lagrangian approach}\\
\vspace*{1.0cm}
{\large
Giulio Cianti$^{1,~}$\footnote{E-mail: giulio.cianti@uniroma1.it} and Enrico Meggiolaro$^{2,~}$\footnote{E-mail: enrico.meggiolaro@unipi.it}
}\\
\vspace*{0.5cm}{\normalsize
$^{1}$ {Dipartimento di Fisica, ``Sapienza'' Universit\`a di Roma, and INFN, Sezione di Roma, Piazzale Aldo Moro 2, I-00185 Roma, Italy}}\\
\vspace*{0.5cm}{\normalsize
$^{2}$ {Dipartimento di Fisica, Universit\`a di Pisa, and INFN, Sezione di Pisa,\\ Largo Pontecorvo 3, I-56127 Pisa, Italy}}\\
\vspace*{2cm}{\large \bf Abstract}
\end{center}

\noindent
In this work, expanding on previous analyses, we employ an effective Lagrangian approach to investigate the mass spectrum of scalar and pseudoscalar mesons at finite temperature, above the (pseudo-)critical temperature $T_c$, in a ``realistic'' $N_f = 2 + 1$ flavor scenario with degenerate \emph{up} and \emph{down} quarks and a heavier \emph{strange} quark: $0 < m_u = m_d \ll m_s$.
The model's predictions are then critically compared with available lattice QCD results (where meson screening masses are extracted from chiral susceptibilities, which correspond to two-point correlation functions of suitable interpolating operators), looking, in particular, for signatures of the breaking of the $U(1)$ axial symmetry above $T_c$.
}
\newpage

\section{Introduction}\label{Introduction}

In the limit of $N_f$ massless quarks (with the most physically relevant cases being $N_f=2$, corresponding to \emph{up} and \emph{down} quarks, and $N_f=3$, corresponding to \emph{up}, \emph{down} and \emph{strange} quarks), a significant feature of \emph{Quantum ChromoDynamics} (QCD) is its \emph{chiral symmetry}. This means that the Lagrangian of the theory is invariant under the transformations inside the chiral group$\,$ $G=U(1)_V\otimes U(1)_A\otimes SU(N_f)_V\otimes SU(N_f)_A$, which extends the well-known $SU(2)_V$ (isospin) symmetry and $SU(3)_V$ (Gell-Mann) symmetry.\\
However, at least at zero temperature, chiral symmetry is not exactly realized, even in the absence of explicit breaking due to quark masses. Indeed, while the $U(1)_V$ (vectorial) symmetry is exactly realized (leading to baryon number conservation), the special unitary part of the group is spontaneously broken down to its vectorial subgroup $SU(N_f)_V$. This is evidenced by the nonzero value of the so-called \emph{chiral condensate} $\langle\Overline[2]{\psi}\psi\rangle \equiv \sum_f^{N_f} \langle\Overline[2]{\psi}_f\psi_f\rangle$ [$\langle \ldots \rangle$ standing for the \emph{vacuum expectation value} (vev) at zero temperature or, more generally, for the thermal average at a finite temperature $T$], which represents an order parameter for the spontaneous breaking of chiral symmetry. The small masses (in the real world with nonzero quark masses) of the $N_f^2-1$ pseudo-Goldstone bosons (the three pions in the case $N_f=2$ and the octet of the pseudoscalar mesons for $N_f=3$) is another strong evidence for the spontaneous breaking of chiral symmetry. Conversely, the $U(1)_A$ (axial) symmetry does not follow the same fate, it being also broken by a quantum anomaly.\\
At high temperatures, the situation changes significantly. Lattice QCD simulations \cite{HotQCD:2018pds,HotQCD:2019xnw} indicate that above a critical temperature, $T_c^{(N_f)}$, thermal fluctuations cause the chiral condensate to vanish, resulting in the restoration of the chiral symmetry $SU(N_f)_V\otimes SU(N_f)_A$. Nevertheless, the $U(1)_A$ symmetry continues to be broken by the quantum anomaly, which, because of the nonzero (even if decreasing) contribution provided by the instanton gas at high temperatures \cite{GPY1981}, is expected to be different from zero also above $T_c^{(N_f)}$. However, the real magnitude of its breaking and its possible \emph{effective} restoration at some temperature $T_{U(1)} \ge T_c^{(N_f)}$ are still important debated questions in hadronic physics (see Ref. \cite{CM2022} for a recent review on this subject).
Studying these symmetry-breaking phenomena and the associated phase transitions is not just a theoretical exercise, but a topic of fundamental importance in both theoretical and experimental particle physics. The behavior of QCD at finite temperatures is particularly relevant in understanding the early universe, where the high-energy conditions present shortly after the Big Bang would have led to a quark-gluon plasma, with chiral symmetry restored.\\
The study of phase transitions in QCD also provides valuable insights into the search for this quark-gluon plasma in heavy-ion collisions.
To model the chiral symmetry breaking and its restoration, \emph{effective field theories} are also commonly used. Although these theories are clearly not fundamental, they should hopefully capture the key features of QCD at low energies. Among the most widely used models there is the \emph{extended linear sigma ($EL_\sigma$) model}, which was initially developed to study chiral dynamics at zero temperature in Refs. \cite{levy_currents_1967,Bardeen:1969ra,Gasiorowicz:1969kn} (see also Refs. \cite{tHooft:1986ooh,hooft_physics_1999}).
Afterwards, this model was applied in a seminal paper by Pisarski and Wilczek \cite{PW1984} and then by others \cite{Lenaghan:2000ey,Roder:2003uz,Butti:2003nu,Basile:2005hw,Pelissetto:2013hqa,Grahl:2013pba} to explore chiral symmetry restoration at finite temperatures, making it a powerful tool for investigating the QCD phase transition.

Partially following the notation introduced in Refs. \cite{MM2013,LM2018,EM2019,EM2023}, the Lagrangian for the extended linear sigma model can be expressed as:
\begin{equation}
	\mathscr{L}_{(EL_\sigma)}(U,U^{\dagger})=\frac{1}{2}\text{Tr}\big[\partial_{\mu}U\partial^{\mu}U^{\dagger}\big]-V(U,U^{\dagger}),
	\label{ELs Lagrangian}
\end{equation}
where the potential $V(U,U^{\dagger})$ is defined as:
\begin{align}\tag*{}
	V(U,U^{\dagger}) &= \frac{1}{4}\lambda^{2}_{\pi}\text{Tr}\big[(UU^{\dagger}-\rho_{\pi}\mathds{1})^2\big]+\frac{1}{4}\lambda'^{2}_{\pi}\text{Tr}\big[UU^{\dagger}\big]^2\\
	&\quad -\frac{B_m}{2\sqrt{2}}\text{Tr}\big[\mathbf{M}(U+U^{\dagger})\big]-k\big[\det U+ \det U^{\dagger}\big] ,
	\label{Potential}
\end{align}
$\mathbf{M} = \diag (m_1, \ldots, m_{N_f})$ being the quark mass matrix.
The $2N_f^2$ degrees of freedom in the model, representing the scalar and pseudoscalar mesonic excitations, are encoded in the complex $N_f\times N_f$ matrix field $U_{ij}$, which can be expressed in terms of the quark fields as
$U_{ij}\sim \Overline[2]{\psi}_j\Big(\frac{\mathds{1}+\gamma^5}{2}\Big)\psi_{i}=\Overline[2]{\psi}_{j,R}\psi_{i,L}$, up to a multiplicative constant.\\
Under $U(N_f)_L\otimes U(N_f)_R$ chiral transformations the quark fields and the mesonic effective field $U$ transform as
\begin{equation}\label{trasfU}
U(N_f)_L\otimes U(N_f)_R:\quad \psi_{L,R}\rightarrow \widetilde{V}_{L,R}\psi_{L,R} ~\Rightarrow~ U\rightarrow \widetilde{V}_L U \widetilde{V}_R^{\dagger} ,
\end{equation}
where $\widetilde{V}_L$ and $\widetilde{V}_R$ are arbitrary $N_f \times N_f$ unitary matrices. Therefore, the first term in the right-hand side of Eq. \eqref{ELs Lagrangian} and the two first terms in the right-hand side of Eq. \eqref{Potential} are invariant under the entire chiral group $U(N_f)_L\otimes U(N_f)_R$, while the last (\emph{anomalous}) interaction term, proportional to the parameter $k$, in the right-hand side of Eq. \eqref{Potential} [and so the entire effective Lagrangian \eqref{ELs Lagrangian} in the \emph{chiral limit} $\mathbf{M}=0$] is invariant under $SU(N_f)_L\otimes SU(N_f)_R \otimes U(1)_V$ but \emph{not} under a $U(1)_A$ transformation:
\begin{equation}\label{U1A}
U(1)_A:\quad \psi_{L,R}\rightarrow e^{\pm i\alpha}\psi_{L,R} ~\Rightarrow~
U \rightarrow e^{i2\alpha} U .
\end{equation}
Therefore, the term proportional to the parameter $k$ in the potential \eqref{Potential} describes (at the level of our effective Lagrangian model) the effects of the $U(1)$ axial anomaly.\\
The temperature dependence in the extended linear sigma model is encoded in the parameters $\rho_{\pi}$, $\lambda_{\pi}$, $\lambda'_{\pi}$, $B_m$, and $k$, all of which are functions of the temperature. In particular, the behavior of the parameter $\rho_{\pi}$ with the temperature plays a central role in determining whether the $SU(N_f)_L\otimes SU(N_f)_R$ chiral symmetry is spontaneously broken or restored. Below the critical temperature $T_c^{(N_f)}$, $\rho_{\pi}$ is positive [$\rho_\pi(T<T_c^{(N_f)})>0$] and the matrix field $U$ has a nonzero vev (i.e., \emph{thermal average}) $\langle U \rangle$, which, in a \emph{mean-field approach}, is just the matrix field configuration $\Overline[2]{U}$ which minimizes the potential $V$. As the temperature rises beyond $T_c^{(N_f)}$, $\rho_{\pi}$ becomes negative [$\rho_\pi(T>T_c^{(N_f)})<0$] and the potential minimum in the chiral limit ($\mathbf{M}=0$) is reached when the vev $\Overline[2]{U}$ is zero, signaling the restoration of chiral symmetry.\footnote{Here we are identifying the critical temperature $T_c^{(N_f)}$ of the chiral phase transition with the temperature $T_{\rho_{\pi}}^{(N_f)}$ at which the parameter $\rho_{\pi}$ is equal to zero: however, this is not always correct. For example, while these temperatures coincide for $N_f=3$, this is not true in the case $N_f=2$, where we have $T_{\rho_{\pi}}^{(2)}<T_c^{(2)}$ (see Refs. \cite{MM2013,EM2019} for a more detailed discussion). The special case $N_f=2+1$, instead, is more intricate and it is discussed in Ref. \cite{EM2023} and recalled in the next section.\label{Foot: T_rho_pi}}\\
For what concerns the parameter $k$ of the anomalous term, on the basis of what we have said at the beginning of this section, it is expected to be always nonzero, even if decreasing in magnitude, at high temperatures: a possible scenario is the one in which the value of $k$, despite being always nonzero, significantly decreases and becomes \emph{negligible} (in some sense) above a certain temperature $T_{U(1)} \ge T_c^{(N_f)}$, thus leading to an \emph{effective} restoration of the $U(1)_A$ symmetry.

The central aim of this paper is to perform a comprehensive analysis of the $EL_\sigma$ model. Specifically, extending the results obtained in previous works \cite{MM2013,EM2023},\footnote{In Ref. \cite{MM2013} a different effective Lagrangian model was used, which, according to the notation already introduced in Refs. \cite{LM2018,EM2019}, is known as the ``interpolating model'', because, in a sense which is explained in those two references, it approximately \emph{interpolates}, when varying the temperature, between the effective Lagrangian model proposed by Witten, Di Vecchia, Veneziano, \emph{et al.} \cite{WDV1,WDV2,WDV3a,WDV3b,WDV3c,WDV3d} at $T=0$ and the above-mentioned ``extended linear sigma model'' for $T>T_c$. Since in this paper we are interested in deriving the results for the mass spectrum only for temperatures above the transition, we have decided to use (for simplicity) the ``extended linear sigma model''.}
we shall study the scalar and pseudoscalar meson mass spectrum above $T_c^{(N_f)}$ in the ``realistic'' case with $N_f=2+1$ light (but not massless) quark flavors. This will be done while maintaining exact $SU(2)_V$ (isospin) symmetry, which means with $0<m_u=m_d\ll m_s$. The predictions of the model will then be critically compared with the results from lattice QCD simulations available in the literature. These lattice QCD simulations typically extract the meson masses from the so-called \emph{chiral susceptibilities} (masses obtained in this way are more specifically called \emph{screening masses}): these susceptibilities correspond to the two-point correlation functions of proper interpolating operators for the meson channels. The restoration of a certain part of the chiral symmetry is expected to manifest through the degeneracy of correlators and \emph{screening masses} in meson channels that are connected by the specific symmetry transformations \cite{Detar:1987kae,Detar:1987hib,Shuryak:1993ee}.

\section{The meson mass spectrum of the $EL_\sigma$ model in the ``ideal'' $N_f=2+1$ case}\label{The ideal Nf=2+1 case}

We first consider the case in which the \emph{up} and \emph{down} quarks are massless, while the \emph{strange} quark remains massive, that is
\begin{equation}
	\mathbf{M}=\begin{bmatrix}
		0 & 0 & 0 \\
		0 & 0 & 0 \\
		0 & 0 & m_{s}
	\end{bmatrix}.
\end{equation}
This specific case, that we shall call the ``ideal'' $N_f=2+1$ case, has been thoroughly examined in a recent study \cite{EM2023}. Here, we present a summary of the key findings from that work.\\
We explicitly write the $2N_f^2$ degrees of freedom inside the field matrix $U$ in terms of the relevant scalar and pseudoscalar mesonic excitations. Above the chiral transition it is convenient to use the following linear parametrization:
\begin{equation}
	U=\frac{1}{\sqrt{2}}\sum_{a=1}^{8}(h_a+i\pi_a)\lambda_a + \frac{1}{\sqrt{3}}(h_0+iS_{\pi})\mathds{1},
	\label{Linear parametrization 1}
\end{equation}
or equivalently:
\begin{equation}
	U=\frac{1}{\sqrt{2}}\sum_{a=1}^{7}(h_a+i\pi_a)\lambda_a + \frac{1}{\sqrt{2}}(\sigma_2+i\eta_2)\mathds{1}_2+(\sigma_s+i\eta_s)\mathds{1}_s,
	\label{Linear parametrization 2}
\end{equation}
where $\mathds{1}$ is the $3\times3$ identity matrix and $\lambda_a$ (with $a\in\{1,...,8\}$) are the eight Gell-Mann matrices, with the normalization $\text{Tr}[\lambda_a\lambda_b]=2\delta_{ab}$. The matrices $\mathds{1}_2$ and $\mathds{1}_s$ are defined as:
\begin{equation}
	\mathds{1}_2\equiv\begin{bmatrix}
		1 & 0 & 0 \\
		0 & 1 & 0 \\
		0 & 0 & 0
	\end{bmatrix},\qquad\mathds{1}_s\equiv\begin{bmatrix}
	0 & 0 & 0 \\
	0 & 0 & 0 \\
	0 & 0 & 1
	\end{bmatrix},
\end{equation}
and the two scalar fields $\sigma_2$ and $\sigma_s$ are the two following linear combinations of $h_8$ and $h_0$:
\begin{equation}\label{sigma_2,sigma_s}
\sigma_2 \equiv \frac{1}{\rad3} h_8 + \sqrt{\frac{2}{3}} h_0 ,\quad
\sigma_s \equiv -\sqrt{\frac{2}{3}} h_8 + \frac{1}{\rad3} h_0 ,
\end{equation}
and, correspondingly, the two pseudoscalar fields $\eta_2$ and $\eta_s$ are the two following linear combinations of $\pi_8$ and $S_\pi$:
\begin{equation}\label{eta_2,eta_s}
\eta_2 \equiv \frac{1}{\rad3} \pi_8 + \sqrt{\frac{2}{3}} S_\pi ,\quad
\eta_s \equiv -\sqrt{\frac{2}{3}} \pi_8 + \frac{1}{\rad3} S_\pi .
\end{equation}
Let us observe that
$h_0,~S_\pi \sim \frac{1}{\rad3} (u\bar{u} + d\bar{d} + s\bar{s})$
are scalar and pseudoscalar $SU(3)$ singlet fields,
and $h_8,~\pi_8 \sim \frac{1}{\sqrt{6}} (u\bar{u} + d\bar{d} - 2 s\bar{s})$
are scalar and pseudoscalar $SU(3)$ octet fields,
while $\sigma_2$, $\eta_2 \sim \frac{1}{\rad2} (u\bar{u} + d\bar{d})$
are scalar and pseudoscalar $SU(2)$ singlet fields
and $\sigma_s$, $\eta_s \sim s\bar{s}$.\\
In other words, the complex matrix field $U$ can be written as $U=H_S+iH_{PS}$, where $H_S$ and $H_{PS}$ are two Hermitian matrices (representing respectively the real part and the imaginary part of $U$) defined as
\begin{equation}
	H_S\equiv\begin{bmatrix}
		\frac{1}{\sqrt{2}}(\sigma_{2}+\delta^{0}) & \delta^{+} & \kappa^{+} \\
		\delta^{-} & \frac{1}{\sqrt{2}}(\sigma_{2}-\delta^{0}) & \kappa^{0} \\
		\kappa^{-} & \bar{\kappa}^{0} & \sigma_{s}
	\end{bmatrix}
\end{equation}
and
\begin{equation}
	H_{PS}\equiv\begin{bmatrix}
		\frac{1}{\sqrt{2}}(\eta_{2}+\pi^{0}) & \pi^{+} & K^{+} \\
		\pi^{-} & \frac{1}{\sqrt{2}}(\eta_{2}-\pi^{0}) & K^{0} \\
		K^{-} & \Bar{K}^{0} & \eta_{s}
	\end{bmatrix}.
\end{equation}
The fields in $H_S$ (linear combinations of $h_a$, $h_0$) are \emph{scalar} ($J^P = 0^+$) mesonic fields, while the fields in $H_{PS}$ (linear combinations of $\pi_a$, $S_\pi$) are \emph{pseudoscalar} ($J^P = 0^-$) mesonic fields.
In particular, the pseudoscalar fields $\pi^0 \equiv \pi_3$, $\pi^\pm \equiv (\pi_1 \mp i\pi_2)/\rad2$ correspond to the \emph{pions} and the pseudoscalar fields $K^\pm \equiv (\pi_4 \mp i\pi_5)/\rad2$, $K^0 \equiv (\pi_6 - i\pi_7)/\rad2$, and $\Bar{K}^0 \equiv (\pi_6 + i\pi_7)/\rad2$ correspond to the \emph{kaons},
while the scalar fields $\delta^0 \equiv h_3$, $\delta^\pm \equiv (h_1 \mp ih_2)/\rad2$ correspond to the scalar partners of pions and the scalar fields $\kappa^\pm \equiv (h_4 \mp ih_5)/\rad2$, $\kappa^0 \equiv (h_6 - ih_7)/\rad2$, and $\bar{\kappa}^0 \equiv (h_6 + ih_7)/\rad2$ correspond to the scalar partners of the kaons.\\
Looking at the Lagrangian of the model without the anomalous term, we can notice that the expression \eqref{ELs Lagrangian} is invariant under the subgroup of chiral transformations $U(2)_L^{(u,d)}\otimes U(2)_R^{(u,d)}$:
$\psi_{L,R}\rightarrow\widetilde{V}_{L,R}^{(u,d)}\psi_{L,R}\implies U\rightarrow\widetilde{V}_{L}^{(u,d)}U\widetilde{V}_{R}^{(u,d)\dagger}$,
where the unitary matrices $\widetilde{V}_L^{(u,d)}$ and $\widetilde{V}_R^{((u,d)}$ mix only the \emph{up} and \emph{down} quarks, leaving the \emph{strange} quark unchanged.\\
We can express these chiral group elements as the composition of a $U(1)_V^{(u,d)} \otimes SU(2)_V^{(u,d)}$ (vectorial) transformation, in which $\widetilde{V}_L^{(u,d)}=\widetilde{V}_R^{(u,d)}\equiv\widetilde{V}^{(u,d)}$, where
\begin{equation}
	\widetilde{V}^{(u,d)}=\begin{bmatrix}
		e^{i\alpha_V} & 0 & 0 \\
		0 & e^{i\alpha_V} & 0 \\
		0 & 0 & 1
	\end{bmatrix}\begin{bmatrix}
		V_{11} & V_{12} & 0 \\
		V_{21} & V_{22} & 0 \\
		0 & 0 & 1
	\end{bmatrix},\quad\text{with}\quad\begin{bmatrix}
		V_{11} & V_{12} \\
		V_{21} & V_{22} \\
	\end{bmatrix}\in SU(2),
\end{equation}
and a $U(1)_A^{(u,d)} \otimes SU(2)_A^{(u,d)}$ (axial) transformation, in which $\widetilde{V}_L^{(u,d)}=(\widetilde{V}_R^{(u,d)})^{\dagger}\equiv\widetilde{A}^{(u,d)}$ where
\begin{equation}
	\widetilde{A}^{(u,d)}=\begin{bmatrix}
		e^{i\alpha_A} & 0 & 0 \\
		0 & e^{i\alpha_A} & 0 \\
		0 & 0 & 1
	\end{bmatrix}\begin{bmatrix}
	A_{11} & A_{12} & 0 \\
	A_{21} & A_{22} & 0 \\
	0 & 0 & 1
	\end{bmatrix},\quad\text{with}\quad\begin{bmatrix}
	A_{11} & A_{12} \\
	A_{21} & A_{22} \\
	\end{bmatrix}\in SU(2).
\end{equation}
Instead, the anomalous interaction term, and the entire effective Lagrangian with it, is invariant under the smaller subgroup $SU(2)_V^{(u,d)}\otimes SU(2)_A^{(u,d)}\otimes U(1)_V^{(u,d)}$, but \emph{not} under $U(1)_A^{(u,d)}$.\\
Under $SU(2)_A^{(u,d)}$ and $U(1)_A^{(u,d)}$ transformations, the meson channels $\sigma_2$, $\eta_2$, $\vec\pi$, and $\vec\delta$ are mixed as follows:
\begin{equation}\label{Degeneracies scheme}
\begin{matrix}
\sigma_2 & \stackrel{U(1)_A^{(u,d)}}{\longleftrightarrow} & \eta_2 \\
SU(2)_A^{(u,d)} \updownarrow & & \updownarrow SU(2)_A^{(u,d)} \\
\vec{\pi} & \stackrel{U(1)_A^{(u,d)}}{\longleftrightarrow} & \vec{\delta}
\end{matrix}
\end{equation}
At zero temperature, the $SU(2)_L^{(u,d)}\otimes SU(2)_R^{(u,d)}$ chiral symmetry is spontaneously broken. However, as temperature rises above the critical value $T_c\equiv T_c^{(2)}$, this symmetry is expected to be restored. When this happens, the $\sigma_2$ and $\vec\pi$ channels become degenerate, with identical masses $M_{\sigma_2} = M_\pi$, and the same happens also for the channels $\eta_2$ and $\vec\delta$. On the other hand, fields related only by $U(1)_A^{(u,d)}$ transformations should retain different masses as long as $k\neq0$.
However, an \emph{effective restoration} of the $U(1)^{(u,d)}$ axial symmetry (i.e., $k \simeq 0$ above $T_c$) should imply that $\sigma_2$ becomes degenerate also with $\eta_2$, and $\vec\pi$ becomes degenerate also with $\vec\delta$,
so that all $\sigma_2$, $\vec\pi$, $\eta_2$, and $\vec\delta$ masses should become the same.

For $T>T_c$, where the $SU(2)_L^{(u,d)}\otimes SU(2)_R^{(u,d)}$ chiral symmetry is expected to be restored, the vev of $U$, that is (in our mean-field approach) the value $\Overline[2]{U}$ for which the potential $V$ is minimum, is given by
\begin{equation}
	\Overline[2]{U}=\begin{bmatrix}
		0 & 0 & 0 \\
		0 & 0 & 0 \\
		0 & 0 & \bar{\sigma}_{s}
	\end{bmatrix},
	\label{stationary point}
\end{equation} 
where $\bar{\sigma}_s$, the vev of the Hermitian scalar field $\sigma_s$, must be a real solution of the following \emph{stationary-point (SP) condition}:
\begin{equation}
	\frac{\partial V}{\partial\sigma_s}\Big|_{SP}=(\lambda^{2}_{\pi}+\lambda'^{2}_{\pi})\bar{\sigma}_{s}^{3}-\lambda^{2}_{\pi}\rho_{\pi}\bar{\sigma}_{s}-\frac{1}{\sqrt{2}}B_m m_{s}=0.
	\label{sp condition}
\end{equation}
All the other first derivatives of the potential with respect to the various scalar and pseudoscalar meson fields come out to be trivially zero when evaluated at the stationary point, which corresponds to the vev $\bar{\sigma}_s$ satisfying the condition \eqref{sp condition} and every other field having vev equal to zero.\\
The quantity $\bar{\sigma}_s$ turns out to be proportional to the \emph{strange}-quark chiral condensate $\langle\Overline[2]{\psi}_s\psi_s\rangle$. To be precise:
\begin{equation}
	\langle\Overline[2]{\psi}_s\psi_s\rangle=\frac{\partial\Overline[2]{V}}{\partial m_s}=-\frac{1}{\sqrt{2}}B_m\bar{\sigma}_s,
\end{equation}
where $\Overline[2]{V}$ is once again the minimum of the potential.\\
By calculating the second derivatives of the potential with respect to the fields and evaluating them at the stationary point \eqref{stationary point}, we can extract the squared masses of the various scalar and pseudoscalar mesonic excitations. The results are:
\begin{align}\tag*{}
	M_{\sigma_{2}}^2=M_{\pi}^2&=-\lambda^{2}_{\pi}\tilde{\rho}_{\pi}-2\tilde{k},\\
	\tag*{}M_{\eta_{2}}^2=M_{\delta}^2&=-\lambda^{2}_{\pi}\tilde{\rho}_{\pi}+2\tilde{k},\\
	M_{K}^2=M_{\kappa}^2=M_{\eta_{s}}^2&=-\lambda^{2}_{\pi}\tilde{\rho}_{\pi}+\lambda^{2}_{\pi}\bar{\sigma}_{s}^2,\label{Spectrum 2+1}\\
	\tag*{}M_{\sigma_{s}}^2&=-\lambda^{2}_{\pi}\tilde{\rho}_{\pi}+\big(3\lambda^{2}_{\pi}+2\lambda'^{2}_{\pi}\big)\bar{\sigma}_{s}^2,
\end{align}
where the squared masses for the $\pi$ and $\delta$ are given by $M_{\pi}^2\equiv M_{\pi_{1}}^2=M_{\pi_{2}}^2=M_{\pi_{3}}^2$ and $M_{\delta}^2\equiv M_{h_{1}}^2=M_{h_{2}}^2=M_{h_{3}}^2$. Similarly, for the kaons and their scalar counterparts, we have $M_{K}^2\equiv M_{\pi_{4}}^2=M_{\pi_{5}}^2=M_{\pi_{6}}^2=M_{\pi_{7}}^2$ and $M_{\kappa}^2\equiv M_{h_{4}}^2=M_{h_{5}}^2=M_{h_{6}}^2=M_{h_{7}}^2$. Moreover, we have introduced the quantities
\begin{equation}\label{rhotilde-kappatilde}
	\tilde{\rho}_{\pi}\equiv\rho_{\pi}-\frac{\lambda'^{2}_{\pi}}{\lambda^{2}_{\pi}}\bar{\sigma}_{s}^2\quad\text{and}\quad\tilde{k}=k\bar{\sigma}_{s},
\end{equation}
in order to write the squared masses more compactly.\\
Of course, the solution outlined above is valid only if it corresponds to a minimum of the potential. This condition is satisfied if all the squared masses in \eqref{Spectrum 2+1} are not negative. A necessary (but not sufficient) condition for this is:
\begin{equation}
	\tilde{\rho}_{\pi}\equiv\rho_{\pi}-\frac{\lambda'^{2}_{\pi}}{\lambda^{2}_{\pi}}\bar{\sigma}_{s}^2\leq0,
	\label{minimum condition}
\end{equation}
since, otherwise, the sum $M_{\pi}^2+M_{\delta}^2=M_{\eta_{2}}^2+M_{\sigma_{2}}^2$ would be negative and violate the positivity condition for squared masses.\\
To understand whether this solution is acceptable or not we can observe, first of all, that Eq. \eqref{sp condition} always admits at least one real and positive solution: in fact, denoting (for brevity) the function of $\bar\sigma_s$ in the left-hand side of Eq. \eqref{sp condition} with $F(\bar\sigma_s)$, one sees that $F(0) = -\frac{1}{\rad 2} B_m m_s < 0$, while $\displaystyle\lim_{x \to +\infty} F(x) = +\infty$. Thus, since $F(x)$ is a continuous function of $x$, there must be at least one positive real root of the function: that is to say, there is at least one positive solution of the stationary-point condition \eqref{sp condition}.\\
Moreover, a real solution of Eq. \eqref{sp condition} \emph{must} be positive, provided that the necessary condition \eqref{minimum condition} (for a solution corresponding to a minimum of the potential) is satisfied. In fact, one immediately sees that $F(x) = \lambda_\pi^2 x^3 - \tilde\rho_\pi(x) x - \frac{1}{\rad 2} B_m m_s \le -\frac{1}{\rad 2} B_m m_s < 0$, if $x<0$ and $\tilde\rho_\pi(x) \equiv \rho_\pi - \frac{\lambda'^{2}_{\pi}}{\lambda^{2}_{\pi}} x^2 \le 0$.
In other words, since the solution of Eq. \eqref{sp condition} which corresponds to the minimum of the potential must necessarily satisfy the condition \eqref{minimum condition}, it must be real and positive: $\bar\sigma_s > 0$.\\
This solution describes (in the case $k>0$) the restoration of the $U(2)_L^{(u,d)}\otimes U(2)_R^{(u,d)}$ chiral symmetry above a certain critical temperature $T_c$, which is defined by the condition:
\begin{equation}
	M_{\pi}^2(T_c)=-\lambda^{2}_{\pi}\rho_{\pi}(T_c)+\lambda'^{2}_{\pi}\bar{\sigma}_{s}^2(T_c)-2k\bar{\sigma}_s(T_c)=0
	\label{critical temperature}
\end{equation}
and, of course, $M_{\pi}^2(T)>0$ for $T>T_c$.\\
From the condition \eqref{critical temperature} it can be seen that the critical temperature $T_c$ would coincide with $T_c^{(3)}$ in the particular limit case in which $\lambda'_{\pi}=k=0$ (or if $m_s=0$). Instead, in the more general case, a nonzero value of the parameter $\lambda'_{\pi}$ has the effect of decreasing the value of the critical temperature $T_c$ with respect to $T_c^{(3)}$, while a nonzero value of the parameter $k$ has the opposite effect of increasing the critical temperature $T_c$ with respect to $T_c^{(3)}$.\\
We now make some comments about the results \eqref{Spectrum 2+1}.
First of all, let us observe that the anomalous term (proportional to the parameter $k$) in the effective Lagrangian, influences the mass spectrum only of the \emph{nonstrange} mesonic excitations, where two $(\frac{1}{2},\frac{1}{2})$ chiral multiplets appear, namely $(\sigma_2,\vec\pi)$ and $(\eta_2,\vec\delta)$, signaling the restoration of the $SU(2)_L^{(u,d)} \otimes SU(2)_R^{(u,d)}$ chiral symmetry, while the squared masses of the nonstrange mesonic excitations belonging to a same $U(1)_A^{(u,d)}$ chiral multiplet, such as $(\sigma_2,\eta_2)$ and $(\vec\pi,\vec\delta)$, are split by the quantity:
\begin{equation}\label{mass-split_1}
	\Delta M_{U(1)_A^{(u,d)}}^2 \equiv M_{\eta_2}^2 - M_{\sigma_2}^2 = M_\delta^2 - M_\pi^2 = 4\tilde{k} = 4k\bar\sigma_s .
\end{equation}
In addition, looking also at the other mesonic excitations containing the \emph{strange} quark flavor, we observe that the $SU(2)_L^{(u,d)} \otimes SU(2)_R^{(u,d)}$ chiral restoration is also manifest in the degeneracy of the (pseudoscalar) \emph{kaons} $K^\pm$ and $K^0$, $\Bar{K}^0$ with their scalar partners $\kappa^\pm$ and $\kappa^0$, $\bar\kappa^0$ (this $K$--$\kappa$ degeneracy was also derived in Ref. \cite{GR2018} analyzing a particular set of QCD Ward Identities).

\section{The meson mass spectrum of the $EL_\sigma$ model in the ``realistic'' $N_f=2+1$ case}

In this section, we will extend the analysis of the effective Lagrangian \eqref{ELs Lagrangian} to the ``realistic'' $N_f=2+1$ case. This means that, in addition to the \emph{strange} quark mass $m_s$, we introduce a small but nonzero mass for the \emph{up} and \emph{down} quarks. We will do this while maintaining exact $SU(2)_V$ (isospin) symmetry, meaning that we consider a common (nonzero) light quark mass $m_l\equiv m_u=m_d$, so that the quark mass matrix $\mathbf{M}$ is now given by
\begin{equation}
	\mathbf{M}=\begin{bmatrix}
		m_{l} & 0 & 0 \\
		0 & m_{l} & 0 \\
		0 & 0 & m_{s}
	\end{bmatrix}.
\end{equation}
The introduction of this light quark mass is particularly important because (as we will see in the next section) it allows us to compare the predictions of the model we are examining with the results of lattice simulations for $T>T_{c}$.\\
The linear parametrization of the matrix field $U$ is unchanged from before: see Eq. \eqref{Linear parametrization 2}.\\
The methods used to derive the results of the model are the same that have been used in the previous section, which will now be described in more detail. First of all, we will find the stationary-point conditions by determining where the first derivatives of the potential $V(U,U^{\dagger})$ vanish. Afterwards, we will expand the potential to the second order around the stationary point and diagonalize the Hessian matrix to extract the mass spectrum, verifying also that the solution that we have found is a real minimum of the potential.

\subsection[Study of the potential and the stationary-point conditions above $T_{c}$]{Study of the potential and the stationary-point conditions above $T_{c}$}\label{Study of the potential}

We shall apply the mean-field approximation, as in the previous section, and search the minimum of the potential \eqref{Potential} among the stationary points, where the first derivatives with respect to the various fields
vanish: this requires solving a system of 18 equations.
Among the possible solutions, we must choose the one corresponding to the minimum of $V(U,U^{\dagger})$, which will be identified by studying the Hessian matrix of the potential. The eigenvalues of this matrix, once proven to be positive, will correspond to the squared masses of the physical states described by the model.

By setting each of these derivatives equal to zero, we obtain the system of equations for the stationary points of the potential $V(U,U^{\dagger})$. However, due to the large number of equations involved, solving this system is not straightforward. Thus, we must make some assumptions to reduce the number of variables and equations that we have to solve, simplifying the resolution of the remaining ones.\\
In particular, we might notice that, thanks to the absence of any $\theta$-term, the Lagrangian \eqref{ELs Lagrangian} is invariant under parity ($P$) and charge ($C$) conjugations. That is because, under these conjugations, the matrix field $U$ changes as follows:
$U\xrightarrow{P}U^{\dagger}$, $U\xrightarrow{C}U^{T}$.
Since we have chosen an Hermitian and symmetric quark mass matrix, $\mathbf{M}$, these two discrete symmetries of the Lagrangian are evident. Therefore, we can reasonably assume that the vacuum state will also be invariant under parity and charge transformations. This assumption can be understood at a deeper level if we consider the quark mass matrix as an external field which explicitly breaks the symmetries of the model. When this term is introduced, the vacuum state ``aligns'' with the external field, exhibiting the same symmetry properties.\\
This assumption ensures that the vev $\Overline[2]{U}$ is both Hermitian and symmetric. Therefore, every pseudoscalar field needs to have a zero vev, while every charged field must have the same vev as its $C$-conjugate. In addition, we have chosen $\mathbf{M}$ to preserve the $SU(2)_V$ (isospin) symmetry. As a result, the vevs of the fields must be invariant under any transformation in the $SU(2)_V^{(u,d)}$ group, meaning that only singlets of this unbroken symmetry group can have a nonzero vev.\\
In addition, the effects of these unbroken continuous and discrete symmetries are significant even in the determination of the form of the Lagrangian expanded around its minimum point. For example, the mixed second derivatives of the potential with respect to a scalar field and a pseudoscalar field, calculated at the minimum point of $V(U,U^{\dagger})$, are zero because of $P$-invariance. This makes the scalar and pseudoscalar sectors of the theory completely decoupled in the mass matrix.\\
The above-reported considerations imply that the vevs of all fields, with the only exceptions of $\sigma_2$ and $\sigma_s$, vanish.
Therefore, the vev of $U$ must be of the following form:
\begin{equation}
	\Overline[2]{U}=\begin{bmatrix}
		\frac{1}{\sqrt{2}}\bar{\sigma}_{2} & 0 & 0 \\
		0 & \frac{1}{\sqrt{2}}\bar{\sigma}_{2} & 0 \\
		0 & 0 & \bar{\sigma}_{s}
	\end{bmatrix},
	\label{stationary point ch3}
\end{equation} 
where $\bar{\sigma}_s$ and $\bar{\sigma}_{2}$, the vevs of the Hermitian fields $\sigma_s$ and $\sigma_2$, must satisfy the following stationary-point conditions:
\begin{equation}
	\begin{cases}
		&(\lambda^{2}_{\pi}+\lambda'^{2}_{\pi})\bar{\sigma}_{s}^{3}+(\lambda'^{2}_{\pi}\bar{\sigma}_{2}^{2}-\lambda^{2}_{\pi}\rho_{\pi})\bar{\sigma}_{s}-\frac{1}{\sqrt{2}}B_m m_{s}-k\bar{\sigma}_{2}^{2}=0 ,\\
		&(\frac{1}{2}\lambda^{2}_{\pi}+\lambda'^{2}_{\pi})\bar{\sigma}_{2}^{3}+(\lambda'^{2}_{\pi}\bar{\sigma}_{s}^{2}-\lambda^{2}_{\pi}\rho_{\pi}-2k\bar{\sigma}_{s})\bar{\sigma}_{2}-B_{m}m_{l}=0 .
	\end{cases}
	\label{SP conditions}
\end{equation}
It is then immediate to say that (reasoning as in Section \ref{The ideal Nf=2+1 case}) $\bar{\sigma}_{2}>0$. First of all, we can treat the first equation of the system as a constraint linking the variables:
\begin{equation}
	\bar{\sigma}_{2}^{2}=\frac{1}{k-\lambda'^{2}_{\pi}\bar{\sigma}_{s}}\Big[(\lambda^{2}_{\pi}+\lambda'^{2}_{\pi})\bar{\sigma}_{s}^{3}-\lambda^{2}_{\pi}\rho_{\pi}\bar{\sigma}_{s}-\frac{1}{\sqrt{2}}B_m m_{s}\Big],
	\label{SP condition 2}
\end{equation}
so $\bar{\sigma}_{2}\rightarrow+\infty$ implies $\bar{\sigma}_{s}\rightarrow\frac{k}{\lambda'^{2}_{\pi}}$. Then, writing the second equation in \eqref{SP conditions} as $G(\bar{\sigma}_2)=0$, we can see that the function $G(x)$ is such that $G(0)<0$ and $\lim\limits_{x\rightarrow+\infty}G(x)=+\infty$, which [for the continuity of $G(x)$] implies that there must be at least one positive real solution of the second equation in \eqref{SP conditions}, $G(\bar{\sigma}_{2})=0$.\\
Once we have found the value of $\bar{\sigma}_2$ that satisfies the second equation in \eqref{SP conditions} with the constraint \eqref{SP condition 2}, we also have a positive value of $\bar{\sigma}_{s}$ that completes the solution of the system. This holds if the numerator and the denominator in the right-hand side of Eq. \eqref{SP condition 2}, both functions with exactly one positive root for $T>T_{c}$, do not vanish simultaneously, i.e.,
$(\lambda_{\pi}^2+\lambda'^{2}_{\pi})\frac{k^3}{\lambda'^{6}_{\pi}}-\lambda_{\pi}^2\rho_{\pi}\frac{k}{\lambda'^{2}_{\pi}}-\frac{1}{\sqrt{2}}B_m m_s\neq0$.
When these two roots are different, this constraint has the interval between them as counterimage of $\mathds{R}^+$. Thus, there is at least one positive solution for the whole system of equations \eqref{SP conditions}.\\
The value of $\bar{\sigma}_{s}$ found in this case should be smaller than the one found in the case $m_{l}=0$ if the denominator in the right-hand side of Eq. \eqref{SP condition 2} is negative (and vice versa). Otherwise, $\bar{\sigma}_{2}^{2}$ would appear to be negative. This means that the value of $\bar{\sigma}_{s}$, which is the root of the numerator of the constraint, gets closer to the root of the denominator when we consider a nonzero $m_l$. This can be proved by showing that the function
$F(x)=(\lambda^{2}_{\pi}+\lambda'^{2}_{\pi})x^{3}-\lambda^{2}_{\pi}\rho_{\pi}x-\frac{1}{\sqrt{2}}B_m m_{s}$
has positive derivative
$F'(x)=3(\lambda^{2}_{\pi}+\lambda'^{2}_{\pi})x^{2}-\lambda^{2}_{\pi}\rho_{\pi}$
around its positive root $x_0$.
In fact, multiplying $F'(x_0)$ by the root $x_0>0$ and using the equation $F(x_0)=0$, we get:
$F'(x_0)x_0=3(\lambda^{2}_{\pi}+\lambda'^{2}_{\pi})x_0^{3}-\lambda^{2}_{\pi}\rho_{\pi}x_0=2(\lambda^{2}_{\pi}+\lambda'^{2}_{\pi})x_0^{3}+\frac{1}{\sqrt{2}}B_m m_{s}>0$.\\
Eq. \eqref{SP condition 2} can also be used to prove (by contradiction) that $\bar{\sigma}_{s}\geq0$ is a necessary condition for the system to have a solution. Indeed, the right-hand side of the equation would have a positive denominator and a negative numerator for $\bar{\sigma}_{s}<0$, disobeying to the stationary-point conditions.\\
Similarly to the $N_f=2$ and the ideal $N_f=2+1$ cases, the vevs $\bar{\sigma}_2$ and $\bar{\sigma}_s$ can be related to the values of the \emph{light} and \emph{strange} chiral condensates as:
\begin{equation}
	\langle\Overline[2]{\psi}_u\psi_u\rangle+\langle\Overline[2]{\psi}_d\psi_d\rangle=\frac{\partial\Overline[2]{V}}{\partial m_l}=-B_m\bar{\sigma}_2,\qquad \langle\Overline[2]{\psi}_s\psi_s\rangle=\frac{\partial\Overline[2]{V}}{\partial m_s}=-\frac{1}{\sqrt{2}}B_m \bar{\sigma}_s,
	\label{Chiral condensates and vevs}
\end{equation}
with $\Overline[2]{V}$ the minimum of the potential.\\
These relations are important because the chiral condensates are quantities that can be computed in lattice simulations. Thus, the ratio between these lattice data would give us the ratio between the vevs of our fields.

\subsection{The squared mass matrix and the spectrum of the theory}\label{The squared mass matrix and the spectrum of the theory}

After these considerations, we can explicitly calculate the squared mass matrix of our potential at the stationary point \eqref{SP conditions}, i.e., the Hessian matrix around the stationary point itself. We will then analyze the sign of its eigenvalues to ensure that the considered stationary point is a minimum of the potential $V(U,U^{\dagger})$.\\
Recalling the observations made about the invariance of the theory under parity, charge conjugation, and $SU(2)_V^{(u,d)}$ transformations, we already know that every mixed derivative of the potential with respect to a scalar field and a pseudoscalar field vanishes at the stationary point. Therefore, to diagonalize the squared mass matrix, we study the scalar and pseudoscalar sectors separately.
Below we present only the second derivatives that are nonzero when evaluated at the stationary point:
\begin{equation}
	\frac{\partial^2 V}{\partial(\delta^{0})^2}=-\lambda^{2}_{\pi}\rho_{\pi}+\Big(\frac{3}{2}\lambda^{2}_{\pi}+\lambda'^{2}_{\pi}\Big)\bar{\sigma}_{2}^2+\lambda'^{2}_{\pi}\bar{\sigma}_{s}^2+2k\bar{\sigma}_{s};
\end{equation}
\begin{equation}
	\frac{\partial^2 V}{\partial(\pi^{0})^2}=-\lambda^{2}_{\pi}\rho_{\pi}+\Big(\frac{1}{2}\lambda^{2}_{\pi}+\lambda'^{2}_{\pi}\Big)\bar{\sigma}_{2}^2+\lambda'^{2}_{\pi}\bar{\sigma}_{s}^2-2k\bar{\sigma}_{s};
\end{equation}
\begin{equation}
	\frac{\partial^2 V}{\partial(\sigma_{2})^2}=-\lambda^{2}_{\pi}\rho_{\pi}+\Big(\frac{3}{2}\lambda^{2}_{\pi}+3\lambda'^{2}_{\pi}\Big)\bar{\sigma}_{2}^2+\lambda'^{2}_{\pi}\bar{\sigma}_{s}^2-2k\bar{\sigma}_{s};
\end{equation}
\begin{equation}
	\frac{\partial^2 V}{\partial(\eta_{2})^2}=-\lambda^{2}_{\pi}\rho_{\pi}+\Big(\frac{1}{2}\lambda^{2}_{\pi}+\lambda'^{2}_{\pi}\Big)\bar{\sigma}_{2}^2+\lambda'^{2}_{\pi}\bar{\sigma}_{s}^2+2k\bar{\sigma}_{s};
\end{equation}
\begin{equation}
	\frac{\partial^2 V}{\partial \delta^{-}\partial\delta^{+}}=-\lambda^{2}_{\pi}\rho_{\pi}+\Big(\frac{3}{2}\lambda^{2}_{\pi}+\lambda'^{2}_{\pi}\Big)\bar{\sigma}_{2}^2+\lambda'^{2}_{\pi}\bar{\sigma}_{s}^2+2k\bar{\sigma}_{s};
\end{equation}
\begin{equation}
	\frac{\partial^2 V}{\partial \pi^{-}\partial\pi^{+}}=-\lambda^{2}_{\pi}\rho_{\pi}+\Big(\frac{1}{2}\lambda^{2}_{\pi}+\lambda'^{2}_{\pi}\Big)\bar{\sigma}_{2}^2+\lambda'^{2}_{\pi}\bar{\sigma}_{s}^2-2k\bar{\sigma}_{s};
\end{equation}
\begin{equation}
	\frac{\partial^2 V}{\partial \bar{\kappa}^{0}\partial\kappa^{0}}=\lambda^{2}_{\pi}\Big(\frac{1}{2}\bar{\sigma}_{2}^{2}+\frac{1}{\sqrt{2}}\bar{\sigma}_{2}\bar{\sigma}_{s}+\bar{\sigma}_{s}^{2}\Big)+\lambda'^{2}_{\pi}\big(\bar{\sigma}_{2}^{2}+\bar{\sigma}_{s}^2\big)-\lambda^{2}_{\pi}\rho_{\pi}+\sqrt{2}k\bar{\sigma}_{2};
\end{equation}
\begin{equation}
	\frac{\partial^2 V}{\partial\Bar{K}^{0}\partial K^{0}}=\lambda^{2}_{\pi}\Big(\frac{1}{2}\bar{\sigma}_{2}^{2}-\frac{1}{\sqrt{2}}\bar{\sigma}_{2}\bar{\sigma}_{s}+\bar{\sigma}_{s}^{2}\Big)+\lambda'^{2}_{\pi}\big(\bar{\sigma}_{2}^{2}+\bar{\sigma}_{s}^2\big)-\lambda^{2}_{\pi}\rho_{\pi}-\sqrt{2}k\bar{\sigma}_{2};
\end{equation}
\begin{equation}
	\frac{\partial^2 V}{\partial \kappa^{-}\partial\kappa^{+}}=\lambda^{2}_{\pi}\Big(\frac{1}{2}\bar{\sigma}_{2}^{2}+\frac{1}{\sqrt{2}}\bar{\sigma}_{2}\bar{\sigma}_{s}+\bar{\sigma}_{s}^{2}\Big)+\lambda'^{2}_{\pi}\big(\bar{\sigma}_{2}^{2}+\bar{\sigma}_{s}^2\big)-\lambda^{2}_{\pi}\rho_{\pi}+\sqrt{2}k\bar{\sigma}_{2};
\end{equation}
\begin{equation}
	\frac{\partial^2 V}{\partial K^{-}\partial K^{+}}=\lambda^{2}_{\pi}\Big(\frac{1}{2}\bar{\sigma}_{2}^{2}-\frac{1}{\sqrt{2}}\bar{\sigma}_{2}\bar{\sigma}_{s}+\bar{\sigma}_{s}^{2}\Big)+\lambda'^{2}_{\pi}\big(\bar{\sigma}_{2}^{2}+\bar{\sigma}_{s}^2\big)-\lambda^{2}_{\pi}\rho_{\pi}-\sqrt{2}k\bar{\sigma}_{2};
\end{equation}
\begin{equation}
	\frac{\partial^2 V}{\partial\sigma_{2}\partial\sigma_{s}}=2\lambda'^{2}_{\pi}\bar{\sigma}_{s}\bar{\sigma}_{2}-2k\bar{\sigma}_{2};
\end{equation}
\begin{equation}
	\frac{\partial^2 V}{\partial(\sigma_{s})^2}=3(\lambda^{2}_{\pi}+\lambda'^{2}_{\pi})\bar{\sigma}_{s}^{2}+\lambda'^{2}_{\pi}\bar{\sigma}_{2}^2-\lambda^{2}_{\pi}\rho_{\pi};
\end{equation}
\begin{equation}
	\frac{\partial^2 V}{\partial\eta_{2}\partial\eta_{s}}=2k\bar{\sigma}_{2};
\end{equation}
\begin{equation}
	\frac{\partial^2 V}{\partial(\eta_{s})^2}=(\lambda^{2}_{\pi}+\lambda'^{2}_{\pi})\bar{\sigma}_{s}^{2}+\lambda'^{2}_{\pi}\bar{\sigma}_{2}^2-\lambda^{2}_{\pi}\rho_{\pi}.
\end{equation}
We then organize each result into scalar and pseudoscalar squared mass matrices. In order to do so, we define the nine-component scalar and pseudoscalar fields:
\begin{equation}
	S=\begin{pmatrix}
		\delta^{0} \\
		\sigma_{2} \\
		\delta^{-} \\
		\delta^{+} \\
		\kappa^{-} \\
		\bar{\kappa}^0 \\
		\kappa^{+} \\
		\kappa^{0} \\
		\sigma_{s} \\
	\end{pmatrix},\quad PS=\begin{pmatrix}
	\pi^{0} \\
	\eta_{2} \\
	\pi^{-} \\
	\pi^{+} \\
	K^{-} \\
	\Bar{K}^0 \\
	K^{+} \\
	K^{0} \\
	\eta_{s} \\
	\end{pmatrix}.
\end{equation}
Thus, the resulting squared mass matrices are:
\begin{equation}
	\mathcal{M}_{S}^2=\begin{bmatrix}
		M_{\delta}^2 & 0 & 0 & 0 & 0 & 0 & 0 & 0 & 0 \\
		0 & M_{\sigma_{2},\sigma_{2}}^2 & 0 & 0 & 0 & 0 & 0 & 0 & M_{\sigma_{2},\sigma_{s}}^2 \\
		0 & 0 & 0 & M_{\delta}^2 & 0 & 0 & 0 & 0 & 0 \\
		0 & 0 & M_{\delta}^2 & 0 & 0 & 0 & 0 & 0 & 0 \\
		0 & 0 & 0 & 0 & 0 & 0 & M_{\kappa}^2 & 0 & 0 \\
		0 & 0 & 0 & 0 & 0 & 0 & 0 & M_{\kappa}^2 & 0 \\
		0 & 0 & 0 & 0 & M_{\kappa}^2 & 0 & 0 & 0 & 0 \\
		0 & 0 & 0 & 0 & 0 & M_{\kappa}^2 & 0 & 0 & 0 \\
		0 & M_{\sigma_{s},\sigma_{2}}^2 & 0 & 0 & 0 & 0 & 0 & 0 & M_{\sigma_{s},\sigma_{s}}^2
	\end{bmatrix},
\end{equation}
\begin{equation}
	\mathcal{M}_{PS}^2=\begin{bmatrix}
		M_{\pi}^2 & 0 & 0 & 0 & 0 & 0 & 0 & 0 & 0 \\
		0 & M_{\eta_{2},\eta_{2}}^2 & 0 & 0 & 0 & 0 & 0 & 0 & M_{\eta_{2},\eta_{s}}^2 \\
		0 & 0 & 0 & M_{\pi}^2 & 0 & 0 & 0 & 0 & 0 \\
		0 & 0 & M_{\pi}^2 & 0 & 0 & 0 & 0 & 0 & 0 \\
		0 & 0 & 0 & 0 & 0 & 0 & M_{K}^2 & 0 & 0 \\
		0 & 0 & 0 & 0 & 0 & 0 & 0 & M_{K}^2 & 0 \\
		0 & 0 & 0 & 0 & M_{K}^2 & 0 & 0 & 0 & 0 \\
		0 & 0 & 0 & 0 & 0 & M_{K}^2 & 0 & 0 & 0 \\
		0 & M_{\eta_{s},\eta_{2}}^2 & 0 & 0 & 0 & 0 & 0 & 0 & M_{\eta_{s},\eta_{s}}^2 
	\end{bmatrix}.
\end{equation}
The nonzero elements of these matrices can be written in terms of the model parameters and the vevs of the fields as:
\begin{align}\tag*{}
	&M_{\delta}^2=-\lambda^{2}_{\pi}\tilde{\rho}_{\pi}+\frac{3}{2}\lambda^{2}_{\pi}\bar{\sigma}_{2}^{2}+2\tilde{k};\\
	\tag*{}&M_{\pi}^2=-\lambda^{2}_{\pi}\tilde{\rho}_{\pi}+\frac{1}{2}\lambda^{2}_{\pi}\bar{\sigma}_{2}^{2}-2\tilde{k};\\
	\label{Mass matrix elements} &M_{\kappa}^2=\lambda^{2}_{\pi}\Big(\frac{1}{2}\bar{\sigma}_{2}^{2}+\frac{1}{\sqrt{2}}\bar{\sigma}_{2}\bar{\sigma}_{s}+\bar{\sigma}_{s}^{2}\Big)-\lambda^{2}_{\pi}\tilde{\rho}_{\pi}+\sqrt{2}\hat{k};\\
	\tag*{}&M_{K}^2=\lambda^{2}_{\pi}\Big(\frac{1}{2}\bar{\sigma}_{2}^{2}-\frac{1}{\sqrt{2}}\bar{\sigma}_{2}\bar{\sigma}_{s}+\bar{\sigma}_{s}^{2}\Big)-\lambda^{2}_{\pi}\tilde{\rho}_{\pi}-\sqrt{2}\hat{k};\\
	\tag*{}&M_{\sigma_{2},\sigma_{2}}^2=-\lambda^{2}_{\pi}\tilde{\rho}_{\pi}+\Big(\frac{3}{2}\lambda^{2}_{\pi}+2\lambda'^{2}_{\pi}\Big)\bar{\sigma}_{2}^{2}-2\tilde{k};\\
	\tag*{}&M_{\sigma_{s},\sigma_{s}}^2=-\lambda^{2}_{\pi}\tilde{\rho}_{\pi}+\big(3\lambda^{2}_{\pi}+2\lambda'^{2}_{\pi}\big)\bar{\sigma}_{s}^2;\\
	\tag*{}&M_{\sigma_{2},\sigma_s}^2=M_{\sigma_{s},\sigma_{2}}^2=2\lambda'^{2}_{\pi}\bar{\sigma}_{s}\bar{\sigma}_{2}-2\hat{k};\\
	\tag*{}&M_{\eta_{2},\eta_{2}}^2=-\lambda^{2}_{\pi}\tilde{\rho}_{\pi}+\frac{1}{2}\lambda^{2}_{\pi}\bar{\sigma}_{2}^{2}+2\tilde{k};\\
	\tag*{}&M_{\eta_{s},\eta_{s}}^2=-\lambda^{2}_{\pi}\tilde{\rho}_{\pi}+\lambda^{2}_{\pi}\bar{\sigma}_{s}^2;\\
	\tag*{}&M_{\eta_{2},\eta_{s}}^2=M_{\eta_{s},\eta_{2}}^2=2\hat{k},
\end{align}
where, as done in Section \ref{The ideal Nf=2+1 case}, we define the squared masses for the $\pi$ and $\delta$ as follows: $M_{\pi}^2\equiv M_{\pi_{1}}^2=M_{\pi_{2}}^2=M_{\pi_{3}}^2$ and $M_{\delta}^2\equiv M_{h_{1}}^2=M_{h_{2}}^2=M_{h_{3}}^2$. Similarly, for the kaons and their scalar counterparts, we defined $M_{K}^2\equiv M_{\pi_{4}}^2=M_{\pi_{5}}^2=M_{\pi_{6}}^2=M_{\pi_{7}}^2$ and $M_{\kappa}^2\equiv M_{h_{4}}^2=M_{h_{5}}^2=M_{h_{6}}^2=M_{h_{7}}^2$. In addition, we have defined the following quantities [which generalize those already defined in Section \ref{The ideal Nf=2+1 case}, Eq. \eqref{rhotilde-kappatilde}], to express the squared masses more compactly:
\begin{equation}
	\tilde{\rho}_{\pi}\equiv\rho_{\pi}-\frac{\lambda'^{2}_{\pi}}{\lambda^{2}_{\pi}}(\bar{\sigma}_{s}^2+\bar{\sigma}_{2}^2),\qquad
	\tilde{k}\equiv k\bar{\sigma}_{s},\qquad
	\hat{k}\equiv k\bar{\sigma}_{2}.
\end{equation}
It is clear that introducing a nonzero light quark mass increases the splitting between scalar and pseudoscalar mesons, with these appearing lighter for positive vevs. As expected, the $SU(2)_V$ (isospin) symmetry is still preserved, as long as the masses of light quarks are identical. Lastly, we should notice that in this case the anomaly gives birth to a correction to the masses even for the \emph{strange} mesons.\\
Among the nine couples of fields, the scalar fields $\sigma_2$ and $\sigma_s$ mix together (being $M_{\sigma_2,\sigma_s}^2 = M_{\sigma_s,\sigma_2}^2 \neq 0$) and the same happens for the two pseudoscalar fields $\eta_2$ and $\eta_s$ (being $M_{\eta_2,\eta_s}^2 = M_{\eta_s,\eta_2}^2 \neq 0$): therefore, the corresponding $2 \times 2$ blocks in the squared mass matrices need to be diagonalized in order to find the squared mass eigenvalues (and the corresponding eigenvectors).
By doing this, we get the scalar and pseudoscalar meson mass spectrum of the theory in terms of the Lagrangian parameters $\rho_{\pi}$, $\lambda_{\pi}$, $\lambda_{\pi}'$, and $k$, and the vevs $\bar{\sigma}_2$ and $\bar{\sigma}_s$:
\begin{align}
	M_{\pi}^2&=-\lambda^{2}_{\pi}\tilde{\rho}_{\pi}+\frac{1}{2}\lambda^{2}_{\pi}\bar{\sigma}_{2}^{2}-2\tilde{k};\label{Mp}\\
	M_{\delta}^2&=-\lambda^{2}_{\pi}\tilde{\rho}_{\pi}+\frac{3}{2}\lambda^{2}_{\pi}\bar{\sigma}_{2}^{2}+2\tilde{k};\label{Md}\\
	\tag*{}M_{K}^2&=\lambda^{2}_{\pi}\Big(\frac{1}{2}\bar{\sigma}_{2}^{2}-\frac{1}{\sqrt{2}}\bar{\sigma}_{2}\bar{\sigma}_{s}+\bar{\sigma}_{s}^{2}\Big)-\lambda^{2}_{\pi}\tilde{\rho}_{\pi}-\sqrt{2}\hat{k}\label{MK}\\
	&=\lambda^{2}_{\pi}\Big(\frac{1}{2}\bar{\sigma}_{2}^{2}+\bar{\sigma}_{s}^{2}\Big)-\lambda^{2}_{\pi}\tilde{\rho}_{\pi}-\frac{1}{\sqrt{2}}\big(\lambda^{2}_{\pi}\bar{\sigma}_{s}+2k\big)\bar{\sigma}_{2};\\
	\tag*{}M_{\kappa}^2&=\lambda^{2}_{\pi}\Big(\frac{1}{2}\bar{\sigma}_{2}^{2}+\frac{1}{\sqrt{2}}\bar{\sigma}_{2}\bar{\sigma}_{s}+\bar{\sigma}_{s}^{2}\Big)-\lambda^{2}_{\pi}\tilde{\rho}_{\pi}+\sqrt{2}\hat{k}\label{Mk}\\
	&=\lambda^{2}_{\pi}\Big(\frac{1}{2}\bar{\sigma}_{2}^{2}+\bar{\sigma}_{s}^{2}\Big)-\lambda^{2}_{\pi}\tilde{\rho}_{\pi}+\frac{1}{\sqrt{2}}\big(\lambda^{2}_{\pi}\bar{\sigma}_{s}+2k\big)\bar{\sigma}_{2};\\
	M_{\eta_l}^2&=-\lambda^{2}_{\pi}\tilde{\rho}_{\pi}+\frac{1}{2}\lambda^{2}_{\pi}\bar{\sigma}_{s}^2+\frac{1}{4}\lambda^{2}_{\pi}\bar{\sigma}_{2}^{2}+\tilde{k}-\sqrt{\frac{1}{4}\Big(\frac{1}{2}\lambda^{2}_{\pi}\bar{\sigma}_{2}^{2}-\lambda^{2}_{\pi}\bar{\sigma}_{s}^2+2\tilde{k}\Big)^2+4\hat{k}^2};\label{Mel}\\
	M_{\eta_h}^2&=-\lambda^{2}_{\pi}\tilde{\rho}_{\pi}+\frac{1}{2}\lambda^{2}_{\pi}\bar{\sigma}_{s}^2+\frac{1}{4}\lambda^{2}_{\pi}\bar{\sigma}_{2}^{2}+\tilde{k}+\sqrt{\frac{1}{4}\Big(\frac{1}{2}\lambda^{2}_{\pi}\bar{\sigma}_{2}^{2}-\lambda^{2}_{\pi}\bar{\sigma}_{s}^2+2\tilde{k}\Big)^2+4\hat{k}^2};\label{Meh}
\end{align}
\begin{align}
	\tag*{}M_{\sigma_l}^2=&-\lambda^{2}_{\pi}\tilde{\rho}_{\pi}+\Big(\frac{3}{2}\lambda^{2}_{\pi}+\lambda'^{2}_{\pi}\Big)\bar{\sigma}_{s}^2+\Big(\frac{3}{4}\lambda^{2}_{\pi}+\lambda'^{2}_{\pi}\Big)\bar{\sigma}_{2}^{2}-\tilde{k}\\
	&-\sqrt{\frac{1}{4}\Big[\Big(\frac{3}{2}\lambda^{2}_{\pi}+2\lambda'^{2}_{\pi}\Big)\bar{\sigma}_{2}^{2}-\big(3\lambda^{2}_{\pi}+2\lambda'^{2}_{\pi}\big)\bar{\sigma}_{s}^2-2\tilde{k}\Big]^2+\big(2\lambda'^{2}_{\pi}\bar{\sigma}_{2}\bar{\sigma}_{s}-2\hat{k}\big)^2};\label{Msl}\\
	\tag*{}M_{\sigma_h}^2=&-\lambda^{2}_{\pi}\tilde{\rho}_{\pi}+\Big(\frac{3}{2}\lambda^{2}_{\pi}+\lambda'^{2}_{\pi}\Big)\bar{\sigma}_{s}^2+\Big(\frac{3}{4}\lambda^{2}_{\pi}+\lambda'^{2}_{\pi}\Big)\bar{\sigma}_{2}^{2}-\tilde{k}\\
	&+\sqrt{\frac{1}{4}\Big[\Big(\frac{3}{2}\lambda^{2}_{\pi}+2\lambda'^{2}_{\pi}\Big)\bar{\sigma}_{2}^{2}-\big(3\lambda^{2}_{\pi}+2\lambda'^{2}_{\pi}\big)\bar{\sigma}_{s}^2-2\tilde{k}\Big]^2+\big(2\lambda'^{2}_{\pi}\bar{\sigma}_{2}\bar{\sigma}_{s}-2\hat{k}\big)^2},\label{Msh}
\end{align}
where $\eta_l$ and $\eta_h$ are the pseudoscalar fields that diagonalize the $2 \times 2$ block $M^2_{\eta_i,\eta_j}$ of the squared mass matrix and are given by:
\begin{equation}
	\begin{cases}
		&\eta_h\equiv\cos\theta_{\eta}\eta_{s}+\sin\theta_{\eta}\eta_{2} ,\\
		&\eta_l\equiv-\sin\theta_{\eta}\eta_{s}+\cos\theta_{\eta}\eta_{2} ,
	\end{cases}
\end{equation}
with
\begin{equation}
	\sin2\theta_{\eta}=\frac{4\hat{k}}{\sqrt{\Big(\frac{1}{2}\lambda^{2}_{\pi}\bar{\sigma}_{2}^{2}-\lambda^{2}_{\pi}\bar{\sigma}_{s}^2+2\tilde{k}\Big)^2+16\hat{k}^2}} ,
\end{equation}
while $\sigma_l$ and $\sigma_h$ are the scalar fields that diagonalize the $2 \times 2$ block $M^2_{\sigma_i,\sigma_j}$ of the squared mass matrix and are given by:
\begin{equation}
	\begin{cases}
		&\sigma_h\equiv\cos\theta_{\sigma}\sigma_{s}+\sin\theta_{\sigma}\sigma_{2} ,\\
		&\sigma_l\equiv-\sin\theta_{\sigma}\sigma_{s}+\cos\theta_{\sigma}\sigma_{2} ,
	\end{cases}
\end{equation}
with
\begin{equation}
	\sin2\theta_{\sigma}=\frac{4(\lambda'^{2}_{\pi}\bar{\sigma}_{s}\bar{\sigma}_{2}-\hat{k})}{\sqrt{\Big[\Big(\frac{3}{2}\lambda^{2}_{\pi}+2\lambda'^{2}_{\pi}\Big)\bar{\sigma}_{2}^{2}-\big(3\lambda^{2}_{\pi}+2\lambda'^{2}_{\pi}\big)\bar{\sigma}_{s}^2-2\tilde{k}\Big]^2+16\big(\lambda'^{2}_{\pi}\bar{\sigma}_{2}\bar{\sigma}_{s}-\hat{k}\big)^2}}.
\end{equation}
The computed spectrum correctly reproduces the behavior found in Refs. \cite{MM2013,EM2023} in the various chiral limits: in particular, in the chiral limit $m_l\rightarrow0$ (in which $\bar{\sigma}_2\rightarrow0$) it reproduces the spectrum of the ideal $N_f=2+1$ case, reported in Eq. \eqref{Spectrum 2+1}, and then, taking also the limit $m_s\rightarrow0$ (by virtue of which $\bar{\sigma}_s\rightarrow0$), it reduces to the spectrum of the ideal case $N_f=3$ (in which $m_u=m_d=m_s=0$).\\
In order to prove that this solution really corresponds to a minimum for the potential $V(U,U^{\dagger})$, we first recall that the stationary-point conditions \eqref{SP conditions} admit a solution for $\bar{\sigma}_2>0$ and $\bar{\sigma}_s>0$. Moreover, the latter assumes values only between the solution of the ideal $N_f=2+1$ case (proven to be positive in Section \ref{The ideal Nf=2+1 case}) and $\frac{k}{\lambda'^2_{\pi}}$. Therefore, $\bar{\sigma}_s$ must be positive.\\
Then, it is useful to notice that, using the expression of $M_{\pi}^2$ and $M_{\eta_{s},\eta_{s}}^2$, we can re-express the system of equations in \eqref{SP conditions} as follows:
\begin{align}
	&M_{\eta_s,\eta_s}^2\bar{\sigma}_{s}-\frac{1}{\sqrt{2}}B_m m_s - k\bar{\sigma}_2^2 = 0;
	\label{Simil-GMOR strange equation}\\
	&M_{\pi}^2\bar{\sigma}_{2} - B_m m_l = 0.
	\label{Simil-GMOR equation}
\end{align}
The latter implies that $M_\pi^2 = B_m m_l/\bar{\sigma}_2 \ge 0 \Leftrightarrow \bar{\sigma}_2 \ge 0$, so that $\bar{\sigma}_{2}\geq0$ is a necessary condition for the stationary point to be a minimum. The first one, instead, indicates that the Hessian matrix element $M_{\eta_s,\eta_s}^2 = \big(\frac{1}{\sqrt{2}}B_m m_s + k \bar{\sigma}_2^2\big)/\bar{\sigma}_s$ is positive.\\
Focusing on the $2\times2$ block $\mathcal{M}_{\eta_i\eta_j}^2$ of the pseudoscalar singlets, we note that $M_{\eta_s,\eta_s}^2$ is positive for the reason just mentioned, while $M_{\eta_2,\eta_2}^2=M_{\pi}^2+4k\bar{\sigma}_s>0$. Therefore both the diagonal elements of the block are positive, which implies that the trace is positive as well. Furthermore, computing the determinant of the $2\times2$ block, we get:
\begin{equation}
\det\mathcal{M}_{\eta_i,\eta_j}^2=M_{\eta_s,\eta_s}^2(M_{\pi}^2+4k\bar{\sigma}_s)-4k^2\bar{\sigma}_2^2=M_{\eta_s,\eta_s}^2M_{\pi}^2+2\sqrt{2}k B_m m_s>0,
\end{equation}
where we have substituted $k\bar{\sigma}_{2}^2$ using \eqref{Simil-GMOR strange equation}. The fact that the trace and the determinant of this $2\times2$ block are positive implies that the eigenvalues $M_{\eta_l}^2$ and $M_{\eta_h}^2$ are positive as well.\\
Similar considerations can be used to prove that also the $2\times2$ block $\mathcal{M}_{\sigma_i\sigma_j}^2$ of the scalar singlets has positive eigenvalues $M_{\sigma_l}^2$ and $M_{\sigma_h}^2$.
The heaviest one, $\sigma_{h}$, can be easily shown to be heavier than the pseudoscalar pion $\pi$:
\begin{align}\tag*{}
	M_{\sigma_h}^2-M_{\pi}^2=&\big(\frac{3}{2}\lambda^{2}_{\pi}+\lambda'^{2}_{\pi}\big)\bar{\sigma}_{s}^2+\big(\frac{1}{4}\lambda^{2}_{\pi}+\lambda'^{2}_{\pi}\big)\bar{\sigma}_{2}^{2}+\tilde{k}\\
	&+\sqrt{\frac{1}{4}\Big[\Big(\frac{3}{2}\lambda^{2}_{\pi}+2\lambda'^{2}_{\pi}\Big)\bar{\sigma}_{2}^{2}-\big(3\lambda^{2}_{\pi}+2\lambda'^{2}_{\pi}\big)\bar{\sigma}_{s}^2-2\tilde{k}\Big]^2+\big(2\lambda'^{2}_{\pi}\bar{\sigma}_{2}\bar{\sigma}_{s}-2\hat{k}\big)^2},
\end{align}
which is clearly positive for $\bar{\sigma}_s>0$.
Therefore, it is enough to demonstrate that the determinant of the $2\times2$ block is positive. Performing the computation, we get:
\begin{align}\tag*{}
	\det\mathcal{M}_{\sigma_i,\sigma_j}^2=&\big[M_{\eta_s,\eta_s}^2+2(\lambda_{\pi}^2+{\lambda'}_{\pi}^2)\bar{\sigma}_s^2\big]M_{\pi}^2\\
	&+\big[(\lambda_{\pi}^2+2{\lambda'}_{\pi}^2)M_{\eta_s,\eta_s}^2+2\lambda_{\pi}^2(\lambda_{\pi}^2+3{\lambda'}_{\pi}^2)\bar{\sigma}_s^2+8{\lambda'}_{\pi}^2k\bar{\sigma}_s-4k^2\big]\bar{\sigma}_2^2,
\end{align}
which reduces to the (positive) determinant of the already diagonal scalar block for $\bar{\sigma}_2=0$ and (for continuity) is surely positive for $m_l \ll m_s$, i.e., approaching the chiral limit $m_l \to 0$ (since in this case $\bar{\sigma}_2 \to 0$).\\
Afterwards, we can prove that the eigenvalues $M_{\delta}^2$, $M_{\kappa}^2$, and $M_{K}^2$ are all positive for the solution that we have found. Indeed, the first two are always larger than $M_{\pi}^2$, being
\begin{align}\tag*{}
	&M_{\delta}^2-M_{\pi}^2=\lambda^{2}_{\pi}\bar{\sigma}_{2}^{2}+4\tilde{k}>0;\\
	&M_{\kappa}^2-M_{\pi}^2=\lambda^{2}_{\pi}\Big(\frac{1}{\sqrt{2}}\bar{\sigma}_{2}\bar{\sigma}_{s}+\bar{\sigma}_{s}^{2}\Big)+2\tilde{k}+\sqrt{2}\hat{k}>0.
\label{delta-pion kappa-pion splittings}
\end{align}
For what concerns the pseudoscalar kaon $K$, instead, from the relation
\begin{equation}
	M_{K}^2-M_{\pi}^2=(\lambda^{2}_{\pi}\bar{\sigma}_{s}+2k)\Big(\bar{\sigma}_{s}-\frac{1}{\sqrt{2}}\bar{\sigma}_{2}\Big)
\end{equation}
one sees that $M_K^2 \ge M_\pi^2 \ge 0$ provided that $\bar{\sigma}_{s}\geq\frac{\bar{\sigma}_{2}}{\sqrt{2}}$ (which is surely verified for $m_l \ll m_s$, i.e., approaching the chiral limit $m_l \to 0$, since in this case $\bar{\sigma}_2 \to 0$).\\
On the other hand, from the relation
\begin{align}\tag*{}
	M_{K}^2-M_{\eta_l}^2=&\frac{1}{2}\lambda^{2}_{\pi}\Big(\bar{\sigma}_{s}^2-\sqrt{2}\bar{\sigma}_{2}\bar{\sigma}_{s}+\frac{1}{2}\bar{\sigma}_{2}^{2}\Big)-\tilde{k}-\sqrt{2}\hat{k}\\
	&+\sqrt{\frac{1}{4}\Big(\frac{1}{2}\lambda^{2}_{\pi}\bar{\sigma}_{2}^{2}-\lambda^{2}_{\pi}\bar{\sigma}_{s}^2+2\tilde{k}\Big)^2+4\hat{k}^2}
\end{align}
one finds that $M_K^2 \ge M_{\eta_l}^2 \ge 0$ provided that
\begin{align}\tag*{}
	\frac{\bar{\sigma}_{2}}{\sqrt{2}}\Big(\bar{\sigma}_{s}-\frac{\bar{\sigma}_{2}}{\sqrt{2}}\Big)&\Big(\bar{\sigma}_{s}-\frac{\bar{\sigma}_{2}}{2\sqrt{2}}-\sqrt{\frac{\bar{\sigma}_{2}^2}{8}+\sqrt{2}\frac{k\bar{\sigma}_{2}}{\lambda^{2}_{\pi}}+2\frac{k^2}{\lambda^{4}_{\pi}}}\Big)\\
	\times&\Big(\bar{\sigma}_{s}-\frac{\bar{\sigma}_{2}}{2\sqrt{2}}+\sqrt{\frac{\bar{\sigma}_{2}^2}{8}+\sqrt{2}\frac{k\bar{\sigma}_{2}}{\lambda^{2}_{\pi}}+2\frac{k^2}{\lambda^{4}_{\pi}}}\Big)\geq0,
\end{align}
which is surely verified for $0\leq\bar{\sigma}_{s}\leq\frac{\bar{\sigma}_{2}}{\sqrt{2}}$. Therefore, combining the two above-reported results, we can conclude that $M_K^2 \ge 0$ for any positive value of $\bar{\sigma}_2$ and $\bar{\sigma}_s$.\\
A notable difference between the realistic and the ideal $N_f=2+1$ case is the nonzero splitting between the scalar and pseudoscalar kaon masses, being
\begin{equation}\label{kappa-K splitting}
M_{\kappa}^2 - M_K^2 = \sqrt{2}\lambda_\pi^2 \bar{\sigma}_s \bar{\sigma}_2 + 2\sqrt{2} k \bar{\sigma}_2 > 0 ,
\end{equation}
for any positive values of $\bar{\sigma}_2$, $\bar{\sigma}_s$, and $k$.
This splitting includes both an \emph{anomalous} contribution, proportional to $k$ (but also to $\bar{\sigma}_2$!) and a remnant \emph{non-anomalous} contribution, proportional to both $\bar{\sigma}_s$ and $\bar{\sigma}_2$ (in the ideal $N_f=2+1$ case $\bar{\sigma}_2=0$ and so this splitting vanishes\dots).\\
A similar result holds for the delta-pion splitting, reported in the first Eq. \eqref{delta-pion kappa-pion splittings}, which, however, turns out to be different from zero even in the ideal $N_f=2+1$ case, since in this case the \emph{anomalous} contribution, proportional to $k$, is also proportional to $\bar{\sigma}_s$, but \emph{not} to $\bar{\sigma}_2$.\\
One of the goals of this paper is to try to find evidence for these anomalous and non-anomalous contributions by comparing the results of the model with the corresponding results from lattice QCD: this will be done in the next section.

\section{Comparison with lattice simulation results}\label{Comparison with lattice simulation results}

After having derived the scalar and pseudoscalar meson mass spectrum within the framework of the $EL_\sigma$ model, we shall try to compare these theoretical predictions with numerical data from lattice QCD.
Several lattice computations of the (\emph{screening}) meson masses at finite temperature (for the case $N_f=2$ and also for the case $N_f=2+1$), exist in the literature \cite{Bernard:1996be,Bernard:1996iz,Kogut1998,Chandrasekharan1999,Karsch:1999vy,Vranas:1999dg,Cheng:2010fe,HotQCD:2012,Aoki2012,Cossu2013,Buchoff2014,Bazavov:2019www,Ding2021}.
In particular, the mass spectrum of mesonic excitations containing both the light (\emph{up} and \emph{down}) and the \emph{strange} quark flavors was studied in detail (by means of lattice simulations) in Ref. \cite{Cheng:2010fe} and in the more recent Ref. \cite{Bazavov:2019www}.
These simulations provide a first-principle and nonperturbative approach based on QCD for studying the behavior of hadronic matter at finite temperatures. However, most of these simulations, including the two mentioned above, are not done in the chiral limit. In fact, they follow a so-called \emph{line of constant physics}. This means that, as explained in \cite{Bazavov:2011nk,HotQCD:2014kol}, they first tune (from simulations of the theory at zero temperature) the mass of the \emph{strange} quark to its physical value and then they set the light quark mass $m_{l}$ to be a fraction of $m_{s}$. In particular in Ref. \cite{Bazavov:2019www}, from which we will take the data for the comparison, the value of $m_s$ is fixed by setting (in the theory at zero temperature) the value of the quantity $\sqrt{2 M_K^2 - M_\pi^2}$ (which, according to chiral perturbation theory at $T=0$, is proportional to $\sqrt{m_s}$) to its physical value $686$ MeV, and then the value $m_l={m_s}/{27}$ is chosen around the transition ($T_{c}=156.5\pm1.5\text{ MeV}$ \cite{HotQCD:2018pds}) and the value $m_l={m_s}/{20}$ is chosen for $T\gtrsim172\text{ MeV}$ (these two values corresponding to a zero-temperature mass for the pion of $140\text{ MeV}$ and $160\text{ MeV}$ respectively).

\subsection{Extrapolated value of $k\bar{\sigma}_2$ and $k\bar{\sigma}_s$}\label{Extrapolated value of kbarsigma2 and kbarsigmas}

The results of the $EL_\sigma$ model that we have studied are completely determined by the five parameters $\rho_{\pi}$, $\lambda_{\pi}$, $\lambda'_{\pi}$, $B_{m}$, $k$ and the two quark masses $m_s$ and $m_l$. However the values of these parameters are not known a priori. Therefore, we need at least five independent types of data from lattice simulation to evaluate the consistency of the model.\\
From the lattice results of Ref. \cite{Bazavov:2019www} we can extract six different screening masses, which are obtained analyzing the spatial correlators of proper interpolating operators $O_{\pi^{-}}$, $O_{K^{-}}$ and $O_{\eta_s}$ for the pseudoscalar sector and $O_{\delta^{-}}$, $O_{\kappa^{-}}$ and $O_{\sigma_s}$ for the scalar sector.
However, as we will see, the system of equations provided by these six masses is not closed. On one hand, this means that the computation of more observables is needed to properly compare the model and the simulations; on the other hand, this gives us consistency conditions that can be used to evaluate the applicability of our framework.\\
First, we add to the system of equations the stationary-point conditions \eqref{SP conditions} with the two vevs $\bar{\sigma}_2$ and $\bar{\sigma}_s$ as variables. Next, since their expressions are easier to manipulate, we substitute the masses $M_{\delta}^2$, $M_{K}^2$ and $M_{\kappa}^2$ with the following splittings:
\begin{align}\tag*{}
	&\Delta_{\delta\pi}\equiv M_{\delta}^2-M_{\pi}^2=\lambda_{\pi}^2\bar{\sigma}_2^2+4k\bar{\sigma}_s;\\
	&\Delta_{K\pi}\equiv M_{K}^2-M_{\pi}^2=(\lambda_{\pi}^2\bar{\sigma}_s+2k)\Big(\bar{\sigma}_{s}-\frac{\bar{\sigma}_{2}}{\sqrt{2}}\Big);\label{Splitting}\\
	\tag*{}&\Delta_{\kappa K}\equiv M_{\kappa}^2-M_{K}^2=\sqrt{2}\lambda_{\pi}^2\bar{\sigma}_s\bar{\sigma}_2+2\sqrt{2}k\bar{\sigma}_2.
\end{align}
The system of equations we consider initially is completed by \eqref{SP conditions} and \eqref{Mp}.\\
From the last two equations in \eqref{Splitting}, we derive the following relation between $\bar{\sigma}_{s}$ and $\bar{\sigma}_{2}$:
\begin{equation}
	\bar{\sigma}_{s}=\Big(2\frac{\Delta_{K\pi}}{\Delta_{\kappa K}}+1\Big)\frac{\bar{\sigma}_2}{\sqrt{2}}\equiv m'\frac{\bar{\sigma}_2}{\sqrt{2}},
	\label{Def m1}
\end{equation}
which eliminates from the system $\bar{\sigma}_{s}$ in favor of $\bar{\sigma}_{2}$ and a quantity measured on the lattice. Using this relation, we can rewrite the last splitting as a useful relation to reduce the power of $\bar{\sigma}_2$ in each equation:
\begin{equation}
	\lambda_{\pi}^2\bar{\sigma}_2^2=\frac{\Delta_{\kappa K}}{m'}-2\sqrt{2}\frac{k}{m'}\bar{\sigma}_2.
	\label{Power reducer}
\end{equation}
These relations, combined with the first splitting, allow us to obtain an equation linking $k\bar{\sigma}_2$ with quantities measured on the lattice:
\begin{equation}
	k\bar{\sigma}_2=\frac{1}{2\sqrt{2}}\frac{m'}{(m')^2-1}\Big(\Delta_{\delta\pi}-\frac{\Delta_{\kappa K}}{m'}\Big).
	\label{ks2}
\end{equation}
\begin{figure}[H]
	\centering
	\includegraphics[width=0.7\textwidth]{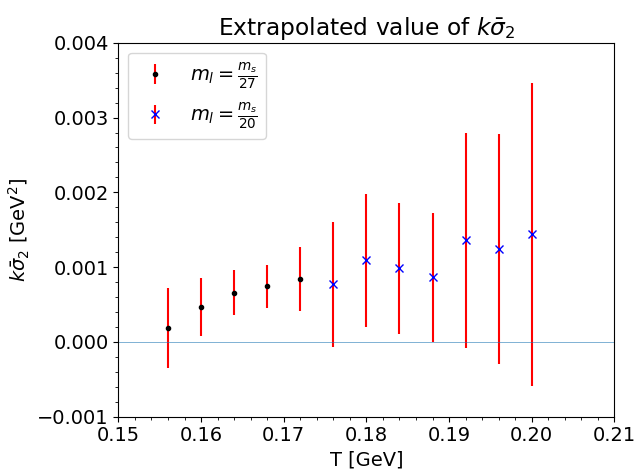}
	\caption[Extrapolated value of $k\bar{\sigma}_2$]{Extrapolated value of $k\bar{\sigma}_2$.}
	\label{Extrapolated value of ks2}
\end{figure}\quad\\
The same thing can obviously be done for $k\bar{\sigma}_s$, which is given by:
\begin{equation}
	k\bar{\sigma}_s=\frac{1}{4}\frac{(m')^2}{(m')^2-1}\Big(\Delta_{\delta\pi}-\frac{\Delta_{\kappa K}}{m'}\Big).
	\label{kss}
\end{equation}
\begin{figure}[H]
	\centering
	\includegraphics[width=0.7\textwidth]{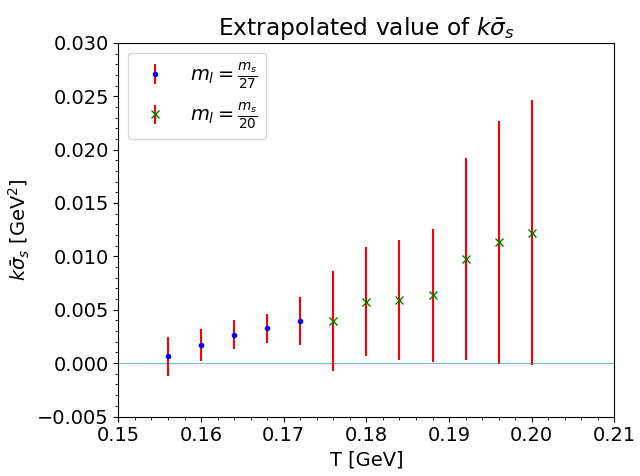}
	\caption[Extrapolated value of $k\bar{\sigma}_s$]{Extrapolated value of $k\bar{\sigma}_s$.}
	\label{Extrapolated value of kss}
\end{figure}\quad\\
The extrapolated values of $k\bar{\sigma}_2$ and $k\bar{\sigma}_s$ are shown in Figures \ref{Extrapolated value of ks2} and \ref{Extrapolated value of kss} respectively:\footnote{The error bars in all the figures reported in this paper have been determined by the propagation of the uncertainties of the screening mass values obtained from lattice QCD in Ref. \cite{Bazavov:2019www}.} 
they are compatible with zero (within the errors) at (i.e., immediately above) $T_c$, but they are positive and incompatible with zero for most of the higher temperatures in the two figures ($T_{c}\leq T\lesssim1.3T_{c}$), thus signaling that the $U(1)_A$ symmetry is \emph{not} effectively restored above $T_{c}$.
Moreover (and quite amazingly!), these two figures (and also the other ones that will be shown below) suggest that the anomalous coupling $k$ of the effective Lagrangian indeed \emph{increases} with the temperature in the range considered (even if it is eventually expected to be proportional to the instanton density \cite{tHooft:1986ooh,hooft_physics_1999,PW1984}, which vanishes asymptotically when $T\to\infty$ \cite{GPY1981}\dots).\\
Once we have inferred the values of $k\bar{\sigma}_2$ and $k\bar{\sigma}_s$, we can use the relations in \eqref{Splitting} to estimate the weight of the \emph{anomalous contribution} to the corresponding mass splitting.\footnote{To clarify the definitions of \emph{anomalous} and \emph{non-anomalous contributions} to a quantity, we define the latter as the value of that quantity in the case $k=0$ and the former as the difference between the quantity and its non-anomalous contribution.}
\begin{figure}[H]
	\centering
	\includegraphics[width=0.49\textwidth]{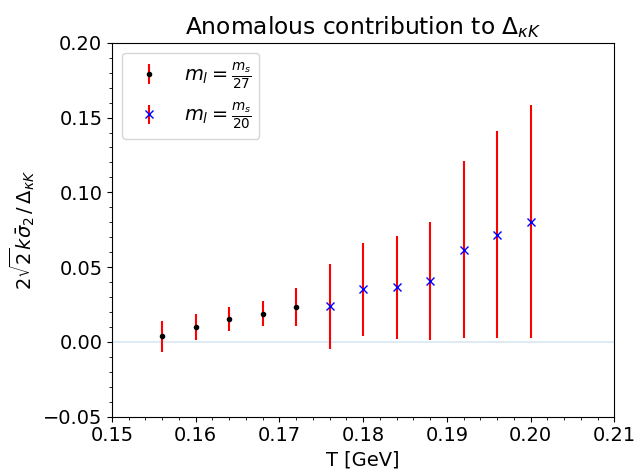}
	\includegraphics[width=0.49\textwidth]{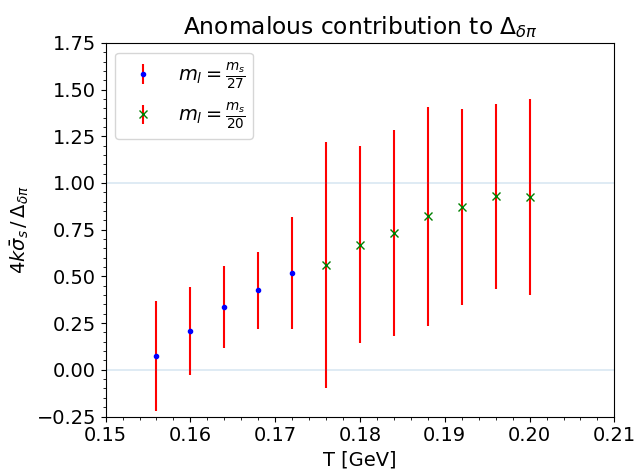}
	\caption[Anomalous contributions to $\Delta_{\kappa K}$ and $\Delta_{\delta\pi}$]{Anomalous contributions to $\Delta_{\kappa K}$ (on the left) and $\Delta_{\delta\pi}$ (on the right).}
	\label{Anomalous contributions}
\end{figure}\quad\\
We observe in Figure \ref{Anomalous contributions} that, while both anomalous contributions grow with the temperature, the one related to $\Delta_{\kappa K}$ remains relatively smaller than the one related to $\Delta_{\delta\pi}$.\\
The evaluation of these relative contributions is important because a common argument for/against the $U(1)_A$ effective restoration at the chiral transition is the zero/nonzero splitting of the masses of the mesons linked by $U(1)_A$ transformations [see Eq. \eqref{Degeneracies scheme}]. However, simulation results obtained with massive quarks also include a non-anomalous contribution to the splitting, which must be considered when studying the $U(1)_A$ effective restoration.

\subsection{Other relevant quantities}\label{Other relevant quantities}

\begin{figure}[H]
	\centering
	\includegraphics[width=0.49\textwidth]{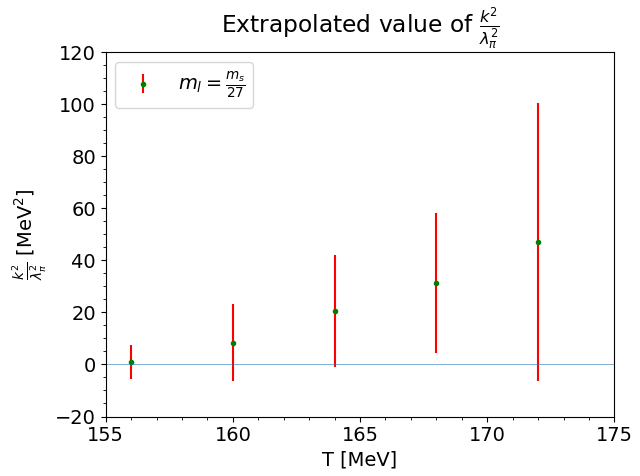}
	\includegraphics[width=0.49\textwidth]{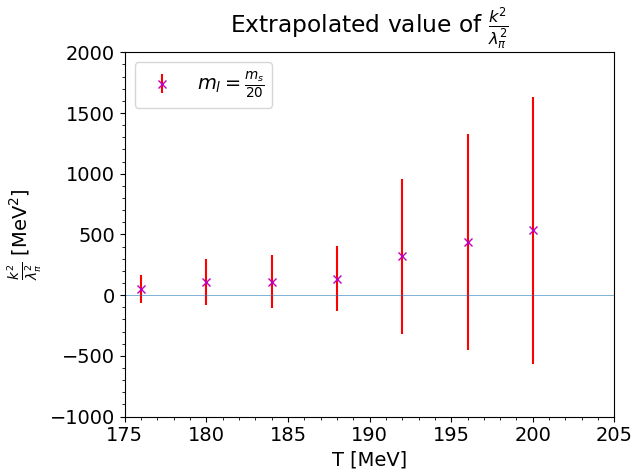}
	\caption[Extrapolated value of $\frac{k^2}{\lambda_{\pi}^2}$]{Extrapolated value of $\frac{k^2}{\lambda_{\pi}^2}$. The two figures have been divided because of the central value variability spanning over three orders of magnitude and the compatibility of the data with zero.}
	\label{Extrapolated value of ic}
\end{figure}

\begin{figure}[H]
	\centering
	\includegraphics[width=0.49\textwidth]{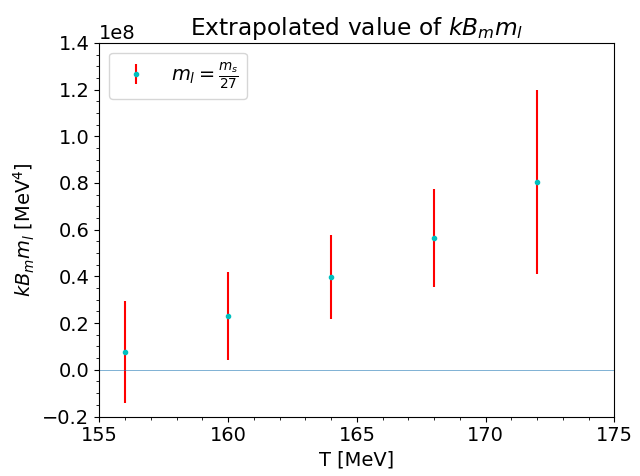}
	\includegraphics[width=0.49\textwidth]{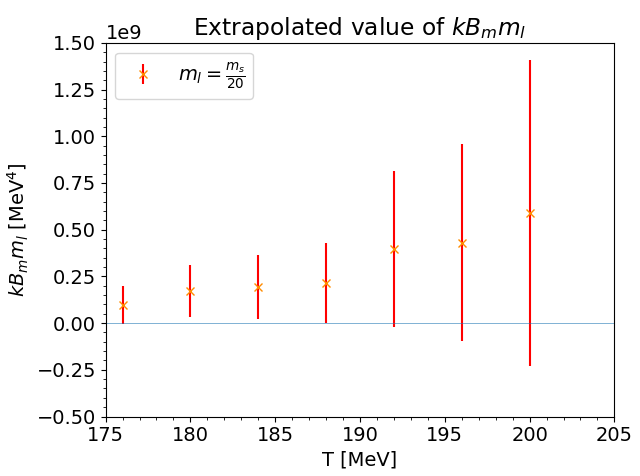}
	\caption[Extrapolated value of $k B_m m_l$]{Extrapolated value of $k B_m m_l$.}
	\label{Extrapolated value of kBmml}
\end{figure}

\begin{figure}[H]
	\centering
	\includegraphics[width=0.7\textwidth]{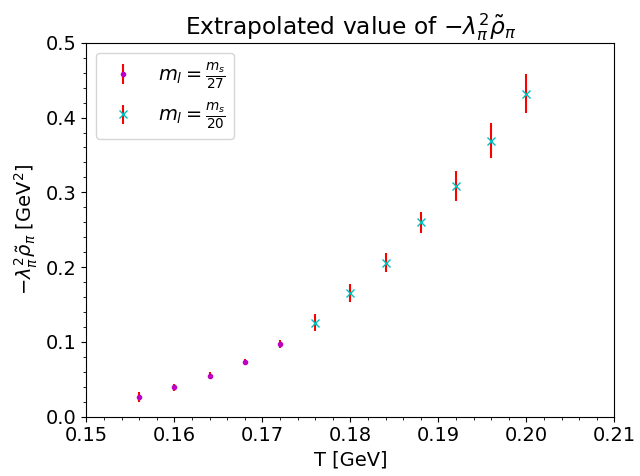}
	\caption[Extrapolated value of $-\lambda_{\pi}^2\tilde{\rho}_{\pi}$]{Extrapolated value of $-\lambda_{\pi}^2\tilde{\rho}_{\pi}$.}
	\label{Extrapolated value of ml2rp1}
\end{figure}

\begin{figure}[H]
	\centering
	\includegraphics[width=0.49\textwidth]{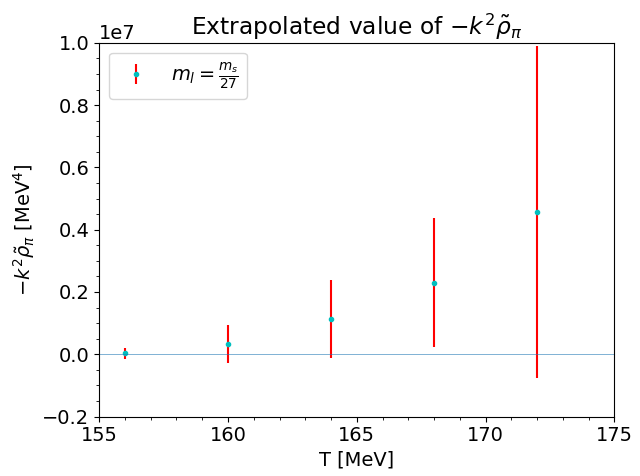}
	\includegraphics[width=0.49\textwidth]{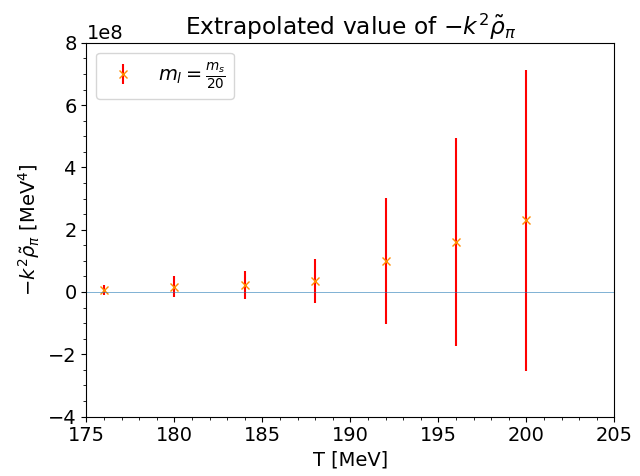}
	\caption[Extrapolated value of $-k^2\tilde{\rho}_{\pi}$]{Extrapolated value of $-k^2\tilde{\rho}_{\pi}$. The two figures have been separated due to the variability of the central value over four orders of magnitude and the compatibility of the data with zero.}
	\label{Extrapolated value of mkrp1}
\end{figure}

Subsequently, to continue with the extrapolations, we can solve the quadratic equation \eqref{Power reducer} and get the only positive solutions: 
\begin{equation}
	\bar{\sigma}_2=\frac{-\sqrt{2}k+\sqrt{2k^2+\lambda_{\pi}^2m'\Delta_{\kappa K}}}{\lambda_{\pi}^2m'}\quad\text{and}\quad\bar{\sigma}_s=\frac{-k+\sqrt{k^2+\frac{1}{2}\lambda_{\pi}^2m'\Delta_{\kappa K}}}{\lambda_{\pi}^2}.
\end{equation}
Next, we can rearrange the first equation and substitute \eqref{ks2} in it to get another quantity which can be inferred from lattice data:
\begin{equation}
	\frac{k^2}{\lambda_{\pi}^2}=\frac{1}{8}\frac{1}{(m')^2-1}\frac{(m'\Delta_{\delta\pi}-\Delta_{\kappa K})^2}{m'\Delta_{\kappa K}-\Delta_{\delta\pi}}.
\end{equation}
The extrapolated values of this quantity are shown in Figure \ref{Extrapolated value of ic}.
\quad\\
Another combination of parameters determinable from lattice data is $k B_m m_l=M_{\pi}^2k\bar{\sigma}_2$. This is nothing else than Eq. \eqref{Simil-GMOR equation} multiplied by $k$.
The extrapolated values of $k B_m m_l$ are shown in Figure \ref{Extrapolated value of kBmml}.
\quad\\
Other useful relations can be found from the first equation of the stationary-point conditions \eqref{SP conditions} and the expression \eqref{Mp} for the squared mass of the pion. Using them we can express in terms of lattice data $-\lambda_{\pi}^2\tilde{\rho}_{\pi}$ and $-k^2\tilde{\rho}_{\pi}$ as follows:
\begin{equation}
	-\lambda_{\pi}^2\tilde{\rho}_{\pi}=\frac{1}{2}\frac{(m')^2+1}{(m')^2-1}\big(M_{\pi}^2+M_{\delta}^2\big)-\frac{M_{K}^2+M_{\kappa}^2}{(m')^2-1}
\end{equation}
and
\begin{equation}
	-k^2\tilde{\rho}_{\pi}=\frac{1}{16}\frac{1}{((m')^2-1)^2}\frac{((m')^2+1)\big(M_{\pi}^2+M_{\delta}^2\big)-2(M_{K}^2+M_{\kappa}^2)}{M_K^2+M_{\kappa}^2-(M_{\pi}^2+M_{\delta}^2)}(m'\Delta_{\delta\pi}-\Delta_{\kappa K})^2.
\end{equation}
The extrapolated values of $-\lambda_\pi^2 \tilde{\rho}_\pi$ and $-k^2 \tilde{\rho}_\pi$ are shown in Figures \ref{Extrapolated value of ml2rp1} and \ref{Extrapolated value of mkrp1} respectively.

\subsection{A consistency condition for the model}\label{A consistency condition for the model}

\begin{figure}[H]
	\centering
	\includegraphics[width=0.7\textwidth]{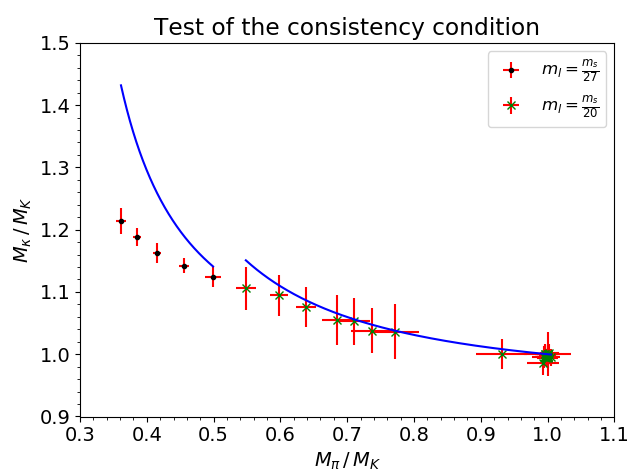}
	\caption[Test of the consistency condition]{Comparison between the lattice data and the prediction \eqref{Consistency condition} of the model. The two curves represent the two different predictions, which come from choosing $m_l=\frac{m_s}{27}$ (on the left) or $m_l=\frac{m_s}{20}$ (on the right). The temperatures corresponding to the displayed data start from $T_{c}$ on the left and increase as we go to the right of the figure.}
	\label{Consistency figure}
\end{figure}

By substituting into the first equation of \eqref{SP conditions} the relations \eqref{Simil-GMOR equation} and \eqref{Def m1}, we obtain the following relation (for $\bar{\sigma}_2\neq0$):
\begin{equation}
	((m')^2-1)\big[m'\lambda_{\pi}^2\bar{\sigma}_2^2+2\sqrt{2}k\bar{\sigma}_2\big]+2M_{\pi}^2\Big(m'-\frac{m_s}{m_l}\Big)=0.
\end{equation}
Then, substituting the relation \eqref{Power reducer} into the squared brackets, we get an expression with no remaining parameters. This indicates that we have established a \emph{consistency condition} for the $EL_\sigma$ model, that is a prediction of the theoretical model concerning the relation between the quark masses and the masses of the scalar and pseudoscalar mesons. This relation can be written as:
\begin{equation}
	\frac{m_s}{m_l}=2\Big(\frac{M_{\kappa}^2-M_{\pi}^2}{M_{\kappa}^2-M_{K}^2}\Big)\frac{M_{K}^2}{M_{\pi}^2}-1,
\end{equation}
or
\begin{equation}
	\frac{M_{\kappa}}{M_{K}}=\sqrt{\frac{\frac{m_s}{m_l}-1}{\big(\frac{m_s}{m_l}+1\big)\big(\frac{M_{\pi}}{M_{K}}\big)^2-2}}\frac{M_{\pi}}{M_{K}}.
	\label{Consistency condition}
\end{equation}
This alternative form of the consistency condition is useful for deriving the behavior of the ratio $M_\kappa/M_K$ in various limit cases.
For example, in the limit of massless \emph{up} and \emph{down} quarks ($m_l \to 0$), with the \emph{strange} quark mass $m_s$ kept fixed to its nonzero value, Eq. \eqref{Consistency condition} reduces to
\begin{equation}
	\lim\limits_{m_l\rightarrow0}\frac{M_{\kappa}}{M_K}=1,
\end{equation}
consistently with the well-known fact that in the \emph{ideal} $N_f=2+1$ case (with $m_u=m_d=0$ and $m_s \neq 0$) the scalar and pseudoscalar kaons become degenerate, as we can see from Eq. \eqref{Spectrum 2+1}.\\
In Figure \ref{Consistency figure} we test this relation using the lattice data from Ref. \cite{Bazavov:2019www}.
\quad\\
We notice that the relation \eqref{Consistency condition} works perfectly for $T\gtrsim172\text{ MeV}$. On the other hand, right above the (pseudo-)critical temperature the expected values and the lattice data differ significantly: this lack of compatibility could be due to various reasons, including, of course, a possible limitation of the effective model used in this paper, or, maybe, an underestimation of the uncertainties.


\section{Summary of the results, comments and conclusions}

In this paper, we have performed an analysis of the \emph{extended linear sigma} ($EL_\sigma$) \emph{model} above the chiral transition. After the seminal paper by Pisarski and Wilczek in 1984 \cite{PW1984}, this model has been often used to study chiral symmetry restoration at finite temperatures. One of the main features of this model is that its Lagrangian contains an \emph{anomalous} term, $k[\det U + \det U^{\dagger}]$, which is invariant under $SU(N_f)_V \otimes SU(N_f)_A \otimes U(1)_V$, but not under $U(1)_A$ transformations \big(while the other terms of the Lagrangian \eqref{ELs Lagrangian}, with the exception of the mass term, are invariant under the entire chiral group $G=U(N_f)_L\otimes U(N_f)_R$\big): this term should reproduce the $U(1)$ axial anomaly of the fundamental theory.\\
Specifically, we have studied this model in the ``realistic'' $N_f=2+1$ case with exact $SU(2)_V$ (isospin) symmetry, considering two degenerate light, but not massless, \emph{up} and \emph{down} quarks and a much heavier \emph{strange} quark (i.e., $0<m_u=m_d\equiv m_l\ll m_s$). We have thus extended the results obtained in previous works \cite{MM2013,EM2023}: in particular, the results of Ref. \cite{EM2023}, concerning the ``ideal'' $N_f=2+1$ case, i.e., the case in which $m_u=m_d=0$ and $m_s\neq0$, have been recalled in Section \ref{The ideal Nf=2+1 case}.\\
Following the same methods used in Refs. \cite{MM2013,EM2023}, in Section 3 we have extracted the meson mass spectrum for the pseudoscalar excitations $\pi^0$, $\pi^{\pm}$, $K^0$, $\Bar{K}^0$, $K^{\pm}$, $\eta_l$, $\eta_h$ and their scalar counterparts $\delta^0$, $\delta^{\pm}$, $\kappa^0$, $\bar{\kappa}^0$, $\kappa^{\pm}$, $\sigma_l$, $\sigma_h$ above the (pseudo-)critical temperature $T_c$. The procedure that we have adopted consists in finding the minimum of the potential by imposing that the vevs of fields satisfy the \emph{stationary-point conditions} \eqref{SP conditions} and that the Hessian matrix has no negative eigenvalue: the square masses of the mesons are just the eigenvalues of this matrix and they are reported in equations \eqref{Mp}--\eqref{Msh} in terms of the parameters of the model $\rho_{\pi}$, $\lambda_{\pi}$, $\lambda_{\pi}'$, $k$, $B_m m_s$, $B_m m_l$, and of the two vevs $\bar{\sigma}_2$ and $\bar{\sigma}_s$. These last two quantities come out to be respectively proportional to the \emph{light} and \emph{strange} chiral condensates, as shown in Eq. \eqref{Chiral condensates and vevs}, and their values are constrained by the stationary-point conditions \eqref{SP conditions}.\\
The computed spectrum correctly reproduces the behavior found in Refs. \cite{MM2013,EM2023} in the various chiral limits: in particular, in the chiral limit $m_l\rightarrow0$ (in which $\bar{\sigma}_2\rightarrow0$) it reproduces the spectrum of the ideal $N_f=2+1$ case, reported in Eq. \eqref{Spectrum 2+1}, and then, taking also the limit $m_s\rightarrow0$ (by virtue of which $\bar{\sigma}_s\rightarrow0$), it reduces to the spectrum of the ideal case $N_f=3$ (in which $m_u=m_d=m_s=0$).\\
A notable difference between the realistic and the ideal $N_f=2+1$ case is the nonzero splitting between the scalar and pseudoscalar kaon masses, reported in Eq. \eqref{kappa-K splitting} and in the third Eq. \eqref{Splitting}:
\begin{equation}
	\Delta_{\kappa K}\equiv M_{\kappa}^2-M_K^2=\sqrt{2}\lambda_{\pi}^2\bar{\sigma}_s\bar{\sigma}_2+2\sqrt{2}k\bar{\sigma}_2.
	\label{Splitting kaoni}
\end{equation}
This splitting includes both an \emph{anomalous} contribution, proportional to $k$ (but also to $\bar{\sigma}_2$!) and a remnant \emph{non-anomalous} contribution, proportional to both $\bar{\sigma}_s$ and $\bar{\sigma}_2$ (we recall that in the ideal $N_f=2+1$ case $\bar{\sigma}_2=0$ and so this splitting vanishes\dots).\\
A similar result holds for the delta-pion splitting, reported in the first Eq. \eqref{delta-pion kappa-pion splittings} and in the first Eq. \eqref{Splitting},
\begin{equation}
	\Delta_{\delta\pi}\equiv M_{\delta}^2-M_{\pi}^2=\lambda_{\pi}^2\bar{\sigma}_2^2+4k\bar{\sigma}_s,
	\label{Splitting pioni}
\end{equation}
which, however, turns out to be different from zero even in the ideal $N_f=2+1$ case, since in this case the \emph{anomalous} contribution, proportional to $k$, is also proportional to $\bar{\sigma}_s$, but \emph{not} to $\bar{\sigma}_2$.\\
One of the goals of this paper was to try to find evidence for these anomalous and non-anomalous contributions by comparing the results of the model with the corresponding results from lattice QCD.\\
Therefore, once obtained the predictions of the model, in Section 4 we have compared them with some recent lattice QCD simulations \cite{Bazavov:2019www}, which were performed in the same realistic $N_f=2+1$ case that we have considered in this paper (so improving a previous study, which was done in Ref. \cite{Cheng:2010fe}). These simulations extracted the so-called \emph{screening masses} of the various mesonic excitations from the long-distance behavior of the \emph{chiral susceptibilities}, which correspond to the two-point spatial correlation functions of appropriate interpolating operators for the meson channels. The comparison between theoretical predictions and lattice results above the (pseudo-)critical temperature $T_{c}=156.5\pm1.5\text{ MeV}$ \cite{HotQCD:2018pds} has given rise to interesting relations involving the parameters of the model and the two vevs $\bar{\sigma}_2$ and $\bar{\sigma}_s$. Putting these relations together, we have been able to express certain combinations of the above-mentioned quantities solely in terms of lattice data.\\
As an example of this analysis, we have derived the expressions \eqref{ks2} for $k\bar{\sigma}_2$ and \eqref{kss} for $k\bar{\sigma}_s$.
The extrapolated values of these quantities have been reported in Figures \ref{Extrapolated value of ks2} and \ref{Extrapolated value of kss} respectively:
they turn out to be compatible with zero (within the errors) at (i.e., immediately above) $T_c$, but they are positive and incompatible with zero for most of the higher temperatures in the two figures ($T_{c}\leq T\lesssim1.3T_{c}$), thus signaling that the $U(1)_A$ symmetry is \emph{not} effectively restored above $T_{c}$.
Moreover (and quite amazingly!), these two figures (and also the other ones that have been shown in Section 4) suggest that the anomalous coupling $k$ of the effective Lagrangian indeed \emph{increases} with the temperature in the range considered (even if it is eventually expected to be proportional to the instanton density \cite{tHooft:1986ooh,hooft_physics_1999,PW1984}, which vanishes asymptotically when $T\to\infty$ \cite{GPY1981}\dots): this is surely an extremely interesting result, which deserves to be investigated with a more detailed and in-depth analysis in the near future.
This result seems indeed to be in agreement with that of Refs. \cite{Fejos2016,Fejos2022}, where, investigating the quantum and thermal fluctuations in the $N_f=2+1$ $EL_\sigma$ model and their effect on the chiral anomaly (using functional renormalization-group techniques), it was found that mesonic fluctuations cause an increase (rather than a decrease) of the $U(1)_A$ anomaly effects, i.e., of the value of the anomalous effective coupling $k$ at finite temperature, above $T_c$.
Also the fact that the anomalous coupling $k$ of the effective Lagrangian turns out to be compatible with zero (within the errors) at (i.e., immediately above) $T_c$ is extremely interesting and surely deserves further investigations, in order to clarify if this is an accidental property, an ``artefact'' of the model, or vice versa if it represents a real physical result. The result that we have found seems indeed to confirm the conjecture proposed in Ref. \cite{Pisarski2024}, according to which the anomalous couplings in any effective model must vanish exactly at the critical temperature $T_c$, implying that the anomalous $U(1)_A$ symmetry is restored precisely at $T_c$ (but, as we have already said, not above $T_c$, where the anomalous couplings are nonzero and are expected to vanish asymptotically when $T \to \infty$, due to the presence of semiclassical instantons).\\
In addition, after having extrapolated the values of $k\bar{\sigma}_2$ and $k\bar{\sigma}_s$, we have compared them with the mass splittings \eqref{Splitting kaoni} and \eqref{Splitting pioni} between scalar and pseudoscalar mesons. Figure \ref{Anomalous contributions} shows that the contribution from the anomalous term is negligible for the $\Delta_{\kappa K}$ mass splitting but highly significant for the $\Delta_{\delta \pi}$ mass splitting.\\
Another interesting relation that we have found by analyzing the model is \eqref{Consistency condition}:
\begin{equation}
	\frac{M_{\kappa}}{M_{K}}=\sqrt{\frac{\frac{m_s}{m_l}-1}{\big(\frac{m_s}{m_l}+1\big)\big(\frac{M_{\pi}}{M_{K}}\big)^2-2}}\frac{M_{\pi}}{M_{K}}.
\end{equation}
By using the lattice data, we have obtained the results shown in Figure \ref{Consistency figure}. The lack of full compatibility between the predictions of the model and the lattice results right above the (pseudo-)critical temperature could be due to various reasons, including, of course, a possible limitation of the effective model used in this paper, or, maybe, an underestimation of the uncertainties (in this respect, see also the comment at the end of these conclusions).

Finally, we want to make a comment about a problem that arises when trying to compare the predictions of the model concerning the masses of the scalar and pseudoscalar $SU(2)_V^{(u,d)}$ singlet mesons, i.e., $\sigma_l$, $\sigma_h$, $\eta_l$, and $\eta_h$, with the lattice results.
In the \emph{ideal} $N_f=2+1$ case (with $m_l=0$), these mesons coincide with $\sigma_2$, $\sigma_s$, $\eta_2$, and $\eta_s$ respectively, and their masses could be (in principle) directly compared with the screening masses obtained by studying the large-distance behavior of the two-point spatial correlators of the interpolating operators
$O_{\sigma_2} \equiv \frac{1}{\sqrt{2}} \big( \Overline[2]{\psi}_u \psi_u + \Overline[2]{\psi}_d \psi_d \big)$,
$O_{\sigma_s} \equiv \Overline[2]{\psi}_s \psi_s$,
$O_{\eta_2} \equiv \frac{i}{\sqrt{2}} \big( \Overline[2]{\psi}_u \gamma_5 \psi_u + \Overline[2]{\psi}_d \gamma_5 \psi_d \big)$,
and $O_{\eta_s} \equiv i\Overline[2]{\psi}_s \gamma_5 \psi_s$.
Instead, in the \emph{realistic} $N_f=2+1$ case (with $m_l \neq 0$), the mass eigenstates $\sigma_l$ and $\sigma_h$ are orthogonal linear combinations of $\sigma_2$ and $\sigma_s$ (and vice versa), and the mass eigenstates $\eta_l$ and $\eta_h$ are orthogonal linear combinations of $\eta_2$ and $\eta_s$ (and vice versa), with a mixing angle that (in both cases) vanishes in the chiral limit $m_l \to 0$.
Ideally, the \emph{asymptotic} large-distance behavior of the two-point spatial correlator of a given interpolating operator $O$ should furnish the screening mass of the lightest state that has a nonzero ``overlap'' with the interpolating operator $O$. This implies that, for example, the asymptotic large-distance behavior of the two-point spatial correlator of the interpolating operator $O_{\eta_{s}}$ should furnish the screening mass of the \emph{lightest} pseudoscalar $SU(2)_V^{(u,d)}$ singlet state $M_{\eta_l}$, even if the operator $O_{\eta_s}$ is expected to have a much larger overlap with the \emph{heaviest} pseudoscalar $SU(2)_V^{(u,d)}$ singlet state $\eta_h$ (at least for $m_l \ll m_s$). 
However, this is only true for lattices with infinite volume: finite-size effects can make this asymptotic limit far from being realized, especially if the mixing angle between $\eta_2$ and $\eta_s$ is small and, therefore, a \emph{pre-asymptotic} large-distance behavior governed by the mass of the heaviest state $\eta_h$ is expected.
In light of these considerations, any comparison with lattice results concerning these $SU(2)_V^{(u,d)}$ singlet mesons should be interpreted with a grain of salt: in this paper we have refrained from an in-depth study of this quite delicate subject and we have decided to leave it to future investigations.

We conclude by observing that it would be surely interesting also to compare the ratio of the two vevs $\bar{\sigma}_2$ and $\bar{\sigma}_s$ in our model, that is to say between the \emph{light} and \emph{strange} chiral condensates, directly with lattice data for these quantities (this has not been possible using only the data from Ref. \cite{Bazavov:2019www}). We leave these comparisons, along with other unexplored aspects (such as the above-mentioned one concerning the masses of the $SU(2)_V^{(u,d)}$ singlet mesons) and potential refinements of the model, to future studies.\\
Future studies should also try to clarify to what extent it is reasonable to compare the various meson masses of our effective model with the corresponding screening masses obtained in lattice QCD. Even if in this work we have tacitly assumed that they approximately coincide, indeed there is no real reason for this to be true, as well as there is no real reason for the screening masses to coincide with the \emph{pole} masses, in particular at high temperatures, where the difference between them can be significant. The assumption made in this work, that these different types of masses approximately coincide, was simply the best working hypothesis we could do at this stage, but it clearly represents a sort of ``systematic uncertainty'' in our procedure: thus, further studies will be surely welcome in order to clarify the real magnitude of this systematic uncertainty (and, if possible, to avoid it).



 



\newpage

\renewcommand{\Large}{\large}

\end{document}